\documentclass{aa}
\usepackage{graphicx}
\usepackage{txfonts}
\usepackage{hyperref}
\usepackage{ulem}
\usepackage{float}
\usepackage{comment}
\hypersetup{
    colorlinks=true,
    linkcolor=blue,
    citecolor=blue,
    filecolor=magenta,      
    urlcolor=cyan,
    pdftitle={Overleaf Example},
    pdfpagemode=FullScreen,
    }

\newcommand{\Lya}{Ly$\rm{\alpha}$\,}

\newcommand{\Hb}{H$\rm{\beta}$ }

\newcommand{\OIII}{[\ion{O}{III}]$_{4960,5008}$ }
\newcommand{\OII}{[\ion{O}{II}] }

\newcommand{\pc}{A2744-PC-z7p9}

\begin{document}

   \title{Before its time: a remarkably evolved protocluster core at $z=7.88$}

   \author{Callum Witten
          \inst{1}    
    \and Pascal A. Oesch\inst{1, 2, 3}
    \and William McClymont\inst{4,5}
    \and Romain A. Meyer \inst{1}
    \and Yoshinobu Fudamoto \inst{6}
    \and Debora Sijacki\inst{4,7}
    \and Nicolas Laporte \inst{8}
    \and Jake S. Bennett \inst{9}
    \and Charlotte Simmonds\inst{4,5}
    \and Emma Giovinazzo \inst{1}
    \and A. Lola Danhaive \inst{4,5}
    \and Laure Ciesla \inst{8}
    \and Cristian Carvajal-Bohorquez \inst{8}
    \and Maxime Trebitsch \inst{10}
    }

   \institute{Department of Astronomy, University of Geneva, Chemin Pegasi 51, 1290 Versoix, Switzerland
   \and
   Cosmic Dawn Center (DAWN), Copenhagen, Denmark
   \and
   Niels Bohr Institute, University of Copenhagen, Jagtvej 128, 2200 Copenhagen, Denmark
   \and
   Kavli Institute for Cosmology, University of Cambridge, Madingley Road, Cambridge CB3 0HA, UK
   \and
   Cavendish Laboratory, University of Cambridge, 19 JJ Thomson Avenue, Cambridge CB3 0HE, UK
   \and 
   Center for Frontier Science, Chiba University, 1-33 Yayoi-cho, Inage-ku, Chiba 263-8522, Japan
   \and
   Institute of Astronomy, University of Cambridge, Madingley Road, Cambridge CB3 0HA, UK
   \and
   Aix Marseille Univ, CNRS, CNES, LAM, Marseille, France
   \and
   Center for Astrophysics, Harvard University, Cambridge, MA 02138, USA
   \and 
   LUX, Observatoire de Paris, Université PSL, Sorbonne Université, CNRS, 75014 Paris, France
   \\
   \email{callum.witten@unige.ch}
             }

   \date{Submitted 8 July 2025}
 
  \abstract 
    {Protoclusters represent the most extreme environments in the very early Universe. They form from large-scale dark matter overdensities, harbouring an overabundance of galaxies fed by large gas reservoirs. Their early and accelerated evolution results in a distinct difference in the properties of galaxies resident in protoclusters versus the field, which is known to be in place by $z\sim 5-6$. We utilise {\it JWST} NIRCam observations of the A2744-z7p9OD protocluster at $z=7.88$ to constrain the properties of resident galaxies. We identify seven new protocluster members, bringing the total number to 23 and the total stellar mass of the protocluster to in excess of $10^{10}\ \rm{M_{\odot}}$. These galaxies are remarkably evolved just 650~Myr after the Big Bang, preferentially showing redder UV-slopes and stronger Balmer breaks than is typical of field galaxies. We use the \texttt{PROSPECTOR} spectral energy distribution fitting code to derive key galaxy properties, finding distinct populations in the core versus the outskirts of the protocluster. The core is largely composed of dusty, massive galaxies which can be characterised as undergoing a synchronised lulling phase, while galaxies in the protocluster outskirts are undergoing recent bursts of star formation. Finally, a strong suppression of the continuum around the Ly$\alpha$-break evidences extreme neutral hydrogen column densities in many resident galaxies ($N_{\rm HI}\gtrsim10^{22.5}\ {\rm cm^{-2}}$). The A2744-z7p9OD system is the most extreme, evolved overdensity yet observed at $z>7$, with higher stellar masses, gas densities, and dust attenuation, revealing the intersection of local environment and high-redshift galaxy formation at their extremes.
    }
   \maketitle

\section{Introduction}
The formation of structure in the early Universe, according to our standard cosmological model, $\Lambda$ cold dark matter \citep[$\Lambda$CDM;][]{peebles_large-scale_1982, blumenthal_formation_1984}, begins with the infall of baryonic matter onto the first dark matter overdensities, eventually cooling sufficiently to collapse and form the first stars and galaxies \citep{jeans_stability_1902, rees_cooling_1977, white_galaxy_1991}. According to hierarchical structure growth, these small-scale structures form first, and through cosmic gas accretion and mergers, larger structures are assembled as a function of cosmic time. While small-scale density perturbations govern galaxy formation, larger-scale perturbations result in loose clusters of dark matter halos. As the Universe continues to expand with time, the gravitational attraction of these clustered regions leads to them expanding at slower rates than average density regions further increasing their overdensity \citep{gunn_infall_1972}, until they eventually become massive virially bound structures. The most extreme of these overdensities, protoclusters, represent some of the rarest objects in the early Universe and are likely to evolve into the most massive structures that we know of today \citep[e.g. the Coma cluster,][]{zwicky_rotverschiebung_1933}. 

The galaxies resident within these most massive dark matter overdensities are believed to be fuelled by considerable gas accretion, and hence are expected to undergo significant and early star formation \citep[e.g.][]{behroozi_average_2013,yajima_forever22_2021,lim_flamingo_2024,morokuma-matsui_forever22_2025, baxter_quantifying_2025}. This means that the properties of protocluster-resident galaxies (PRGs) should diverge from those of field galaxies at a relatively early epoch \citep{overzier_realm_2016}. Moreover, this abundant fuelling of baryonic matter onto these dark matter overdensities, makes them ideal environments for powering the rapid black hole growth required to reproduce the most massive black holes observed at $z\sim6$ \citep{sijacki_growing_2009,lupi_high-redshift_2019,huang_early_2020,trebitsch_obelisk_2021,costa_agn-driven_2022,zhu_formation_2022,costa_host_2024,bennett_growth_2024}.

\begin{figure*}
\centering
\includegraphics[width=1\textwidth]{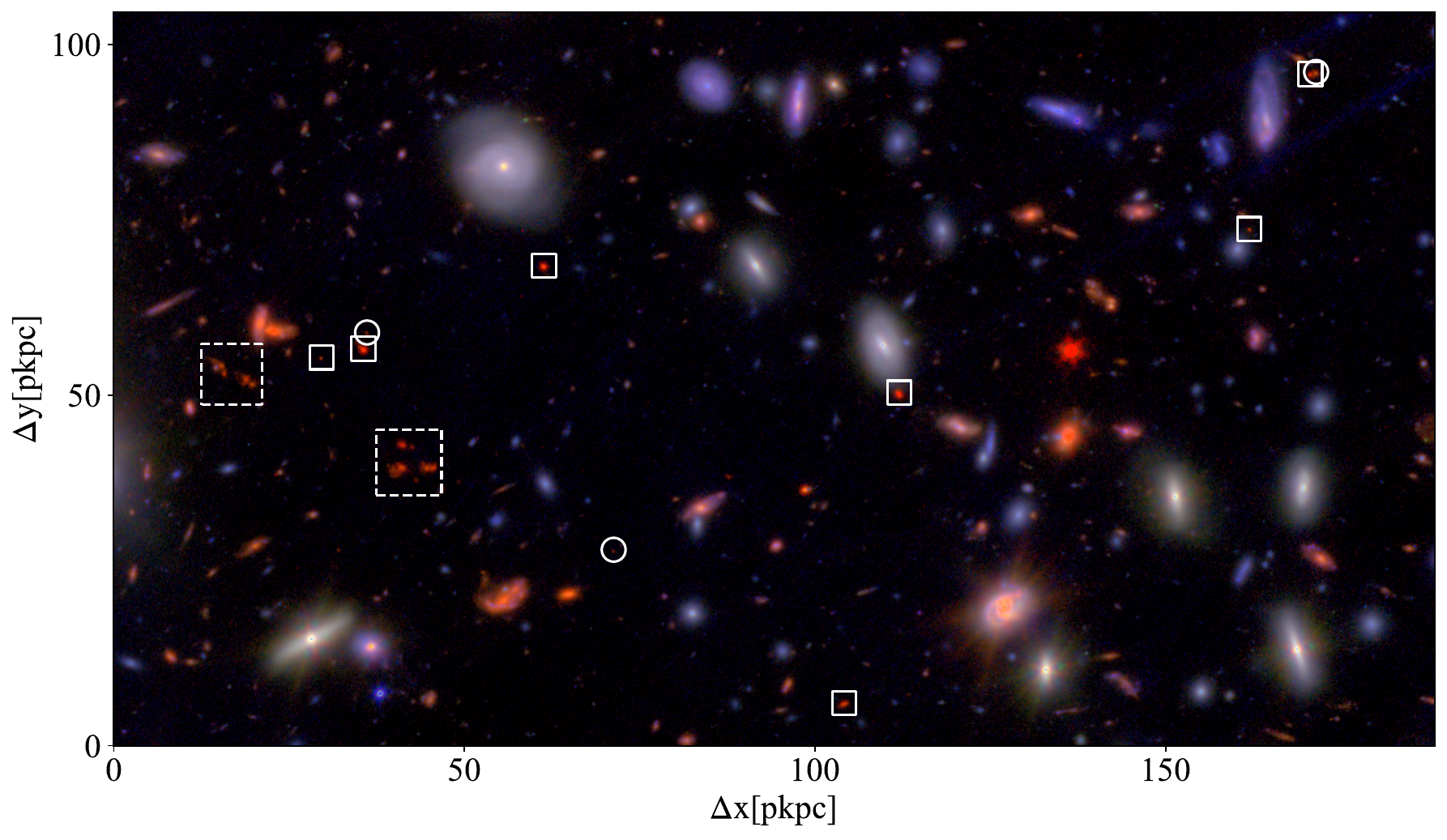}
\caption{An RGB image using the F090W, F277W and F444W filters, with overlaid points indicating the positions of spectroscopically confirmed galaxies (squares, solid line) and photometric candidates (circles). The dashed, large squares indicate the most clustered core regions. The left square includes four PRGs, while the right square contains seven PRGs, these objects compose the ``core'' regions (as defined in Section~\ref{sec:proto_props}). Two of the 23 galaxies are not shown in this figure as they lie at a larger separation than the chosen field-of-view. The axis ticks are intended to indicate the approximate distances between galaxies after correcting for magnification (assuming a constant $\mu = 1.9$, while this does vary across the RGB image, this effect is relatively small).}
\label{fig:RGB}
\end{figure*}

Although the divergence in the properties of PRGs compared to field galaxies is observed out to $z\sim5-6$ \citep[e.g. galaxies with large Balmer breaks and low equivalent width emission lines; ][]{morishita_accelerated_2025,arribas_ga-nifs_2024}, it remains uncertain when this divergence begins. Early evidence of relative dust enrichment and evolved stellar populations can be seen in the UV continuum slopes of galaxies in overdense environments \citep{li_epochs_2025} and indeed their continuum and emission line properties \citep{witten_rising_2025}. However, in order to understand how significant this divergence is, one requires deep imaging across numerous photometric bands to identify a large number of PRGs and accurately estimate their properties. 

Among almost all of the deep {\it JWST} Near-Infrared Camera (NIRCam) imaging campaigns there have been identifications of regions that host significant overdensities of galaxies \citep[e.g.][]{castellano_early_2023,witstok_inside_2024,tang_jwstnirspec_2023,tacchella_jades_2023,scholtz_gn-z11_2024,fudamoto_sapphires_2025, laporte_lensed_2022,herard-demanche_mapping_2025,helton_identification_2024, helton_jwst_2024, brinch_cosmos2020_2023}. However, of those $z>7$ overdensities, very few are sufficiently overdense ($\delta_{\rm gal}\gtrsim10$), within relatively small surface areas ($R\sim 100$~pkpc) and correspondingly small redshift ranges ($\Delta z\sim 0.01$), as to be comparable to what may be expected of a protocluster at $z>7$ \citep[e.g.][]{chiang_galaxy_2017}. Of these, only A2744-z7p9OD (hereafter \pc) has the deep and dense photometric observations necessary to sufficiently characterise the galaxies resident within it from photometry. Moreover, significant previous studies of \pc\ \citep{zheng_young_2014,ishigaki_very_2016,laporte_dust_2017,hashimoto_reionization_2023,morishita_early_2023, morishita_accelerated_2025,witten_rising_2025} have begun to constrain the properties of this protocluster. Most notably, dust detections with the Atacama Large Millimeter Array (ALMA) from \cite{laporte_dust_2017} and \cite{ hashimoto_reionization_2023} and spectroscopic analysis from \cite{witten_rising_2025} indicate that the core of this protocluster hosts at least one relatively evolved galaxy. As such, we utilise recent NIRCam imaging data, discussed in Section~\ref{sec:data}, of this overdense region to identify and characterise protocluster galaxies in Sections~\ref{sec:selection} and~\ref{sec:phot_props}, respectively. We then use the spectral energy distribution (SED)-fitting code \texttt{PROSPECTOR} to further constrain the properties of these galaxies in Section~\ref{sec:sed-fitting}. We then place these results in the context of both the evolution of the protocluster and the distribution of galaxy properties within the protocluster itself in Section~\ref{sec:proto_props}, and finally make our conclusions in Section~\ref{sec:conclusions}.

Throughout this paper all of the galaxy properties that are reported are corrected for magnification. In order to do so, we utilise the $z=8$ magnification map detailed in \cite{furtak_uncovering_2023,price_uncover_2025}. 

\section{Data}
\label{sec:data}

The Abell 2744 lensing field has been subject to a multitude of {\it JWST} observations, making it, to date, the best studied lensing field in terms of near-infrared imaging and spectroscopy. The field has been the subject of Near Infrared Imager and Slitless Spectrograph (NIRISS) imaging and wide field slitless spectroscopy (WFSS) \citep[e.g. GLASS;][]{treu_glass-jwst_2022}, NIRCam imaging with wide bands \citep[UNCOVER;][]{bezanson_jwst_2024} and medium bands \citep[Megascience;][]{suess_medium_2024}, NIRCam WFSS \citep[e.g. ALT;][]{naidu_all_2024}, NIRSpec Spectroscopy \citep[e.g. UNCOVER;][and DDT 2756, PI W. Chen]{bezanson_jwst_2024} and, in the near future, Mid-Infrared Instrument (MIRI) imaging (GO 5578; PI E. Iani). This wealth of data has facilitated the identification of the protocluster core \citep{morishita_early_2023, hashimoto_reionization_2023}, as well as further resident galaxies \citep{roberts-borsani_early_2022, chen_jwst_2024, cameron_nebular_2024} and some initial characterisation of their properties \citep[e.g.][]{morishita_accelerated_2025, witten_rising_2025}. 

We chose to utilise all of the available NIRCam imaging released by the UNCOVER survey \citep{bezanson_jwst_2024} and the MEGASCIENCE survey \citep{suess_medium_2024} to produce $R\sim 15$ spectro-photometry across the wavelength range $0.6\mu{\rm m}\lesssim\lambda\lesssim5\mu{\rm m}$. This resolution facilitates strong constraints on the continuum and emission-line properties of detected galaxies, allowing for unrivalled estimates of star-formation histories (SFHs) and hence stellar masses and galaxy ages.

Although for many of the galaxies we study, spectroscopy is available, and we leverage the spectroscopically confirmed redshifts throughout this work (see Table~\ref{tab:Targets}), these span a range of different instruments which leads to challenges interpreting the results from instrument to instrument. For example, when NIRSpec micro-shutter assembly (MSA) spectroscopy is available, it is often not clear how the emission line flux measured in a small MSA aperture can be related to that from a photometric aperture. While WFSS spectroscopy can resolve this problem, thanks to the lack of any apertures, the available NIRCam WFSS spectroscopy in the A2744 field utilises the F356W, F410M or F460M filter, all of which miss the strongest emission lines (\OIII and ${\rm H\beta}$) at the redshift of the protocluster. As such, we largely avoid spectroscopic analysis in this work (except where this is used for spectroscopic redshift measurements), focusing instead on the deep, relatively high spectral-resolution NIRCam imaging in order to provide the least biased view of this protocluster. 

The NIRCam photometric catalogue that we use to identify photometric candidate resident galaxies is produced following \cite{weibel_galaxy_2024} as described in \cite{naidu_all_2024} for the Abell 2744 cluster. In order to produce robust photometry of PRGs for SED-fitting, we extract the photometry of selected candidate resident galaxies using the \texttt{Photutils} package. We place custom apertures over each candidate, ensuring that we avoid contamination from nearby sources. The motivations for repeating the photometric extraction are that it allows us to use customised apertures and to produce a local background subtraction, which removes potential issues arising from intra-cluster light from the foreground lensing cluster. 

\section{Selection of protocluster-resident galaxies}
\label{sec:selection}

\begin{table*}[t]
\centering
\caption{The properties of PRGs. The following citations are for the spectroscopic redshift: [1] \citealt{hashimoto_reionization_2023}; [2] \citealt{venturi_gas-phase_2024}; [3] \citealt{morishita_accelerated_2025}; [4] \citealt{morishita_early_2023}; [5] this work, based on UNCOVER spectra (see Appendix~\ref{app:spec}); [6] \citealt{naidu_all_2024}. The UV-slopes of s1 and ZD1 could not be established due to the low signal-to-noise of their SEDs. The magnification factor is taken from the $z\sim 8$ magnification map from \cite{furtak_constraining_2022,price_uncover_2025}.}
\label{tab:Targets}
 \begin{tabular}{lccccccc}
  \hline
  Name & RA & Dec. &  $z_{\rm phot}$ & $z_{\rm spec}$ & $M_{\rm UV}$ & $\beta$ & $\mu$ \\
  \hline
  \multicolumn{3}{l}{\bf Spectroscopically confirmed galaxies}&&&&\\
  YD4    & 3.60384 & -30.38224 & $7.91\pm0.04$ & $7.8742 \pm 0.0001$ $^{[1]}$        & $-19.71^{+0.02}_{-0.02}$ & $-1.5 \pm 0.1$ & 1.95 \\
  YD6    & 3.60398 & -30.38227 & $7.95\pm0.03$ & $7.880 \pm 0.001$ $^{[2]}$          & $-18.99^{+0.04}_{-0.04}$ & $-1.9 \pm 0.2$ & 1.95 \\
  YD7-W  & 3.60327 & -30.38224 & $7.89\pm0.05$ & $7.8721 \pm 0.0001 $ $^{[1]}$       & $-19.08^{+0.03}_{-0.03}$ & $-1.3 \pm 0.2$ & 1.96 \\
  s1     & 3.60365 & -30.38192 & $0.06^{+7.88}_{-0.02}$ & $7.8732 \pm 0.0001$ $^{[1]}$   & $-18.00^{+0.10}_{-0.11}$ & NA & 1.94 \\
  YD1    & 3.60383 & -30.38189 & $7.89\pm0.05$ & $7.8778 \pm 0.0001$  $^{[1]}$      & $-19.04^{+0.05}_{-0.05}$ & $-1.9 \pm 0.3$ & 1.94 \\
  ZD3    & 3.60645 & -30.38098 & $7.91\pm0.04$ & $7.8808 \pm 0.0001$  $^{[3]}$       & $-19.47^{+0.03}_{-0.03}$ & $-2.2 \pm 0.1$  & 1.87 \\
  ZD6    & 3.60656 & -30.38090 & $7.91\pm0.04$ & $7.8797 \pm 0.0001$  $^{[3]}$       & $-19.23^{+0.03}_{-0.03}$ & $-2.1 \pm 0.1$ & 1.86 \\
  ZD12-W & 3.60694 & -30.38081 & $7.91\pm0.04$ & $7.8759 \pm 0.0001$  $^{[3]}$       & $-18.98^{+0.05}_{-0.06}$ & $-2.1 \pm 0.3$ & 1.86 \\
  ZD12-E & 3.60699 & -30.38070 & $7.91\pm0.04$ & $7.8782 \pm 0.0001$  $^{[3]}$       & $-19.45^{+0.03}_{-0.03}$ & $-2.5 \pm 0.1$ & 1.86 \\
  ZD2    & 3.60452 & -30.38044 & $7.89\pm0.05$ & $7.8800 \pm 0.0001$  $^{[4]}$       & $-20.36^{+0.02}_{-0.02}$ & $-2.1 \pm 0.1$ & 1.89 \\
  ZD4    & 3.60525 & -30.38058 & $7.81\pm0.08$ & $7.883  \pm 0.001$   $^{[5]}$       & $-18.04^{+0.09}_{-0.10}$ & $-2.1 \pm 0.5$ & 1.87 \\
  z8-2   & 3.60134 & -30.37920 & $7.88\pm0.06$ & $7.8831 \pm 0.0001$  $^{[4]}$       & $-20.04^{+0.02}_{-0.03}$ & $-2.1 \pm 0.1$ & 2.03 \\
  YD8    & 3.59608 & -30.38582 & $7.91\pm0.04$ & $7.8869 \pm 0.0001$  $^{[4]}$       & $-19.25^{+0.03}_{-0.03}$ & $-2.1 \pm 0.1$  & 2.76 \\
  ALT-41799 & 3.59512 & -30.38112 & $7.88\pm0.06$& $7.883 \pm 0.005$  $^{[6]}$       & $-19.87^{+0.02}_{-0.02}$ & $-1.8 \pm 0.1$ & 2.61 \\
  \hline
  \multicolumn{3}{l}{\bf Previously identified candidates}&&&&&\\
  ZD1    & 3.60358 & -30.38243 & $7.9\pm0.2$  &           & $-17.61^{+0.12}_{-0.13}$ & NA & 1.96 \\
  YD7-E  & 3.60339 & -30.38224 & $7.93\pm0.04$ &          & $-19.52^{+0.02}_{-0.02}$ & $-2.1 \pm 0.1$ & 1.96 \\
  \hline
  \multicolumn{3}{l}{\bf Newly identified candidates and confirmed galaxies}&&&&&\\
  PC1& 3.60444 & -30.38022 & $7.88\pm0.04$   &      & $-17.67^{+0.14}_{-0.16}$ & $-2.2 \pm 0.4$  & 1.88 \\
  PC2-E & 3.58790 & -30.37630 & $7.90\pm0.05$  &  $7.8798 \pm 0.001 $ $^{[5]}$ & $-18.26^{+0.05}_{-0.05}$ & $-2.5 \pm 0.3$  & 2.75 \\
  PC2-W & 3.58781 & -30.37628 & $7.93\pm0.03$   &  & $-18.98^{+0.03}_{-0.03}$ & $-2.5 \pm 0.1$ & 2.75 \\
  PC3   & 3.61349 & -30.43473 & $7.88\pm0.09$  &    & $-19.53^{+0.02}_{-0.02}$ & $-2.4 \pm 0.1$ & 1.53 \\
  PC4   & 3.61573 & -30.36839 & $7.89\pm0.05$   &   & $-18.60^{+0.08}_{-0.08}$ & $-2.2 \pm 0.3$ & 1.57 \\
  PC5   & 3.58897 & -30.37864 & $7.91\pm0.04$  &  $7.8709 \pm 0.001$ $^{[5]}$  & $-17.63^{+0.10}_{-0.11}$ & $-2.0 \pm 0.4$ & 2.72 \\
  PC6 & 3.60012 & -30.38350 & $7.91\pm0.09$ & &$-17.35^{+0.17}_{-0.21}$&$-1.6\pm0.5$&2.14\\
  \hline
 \end{tabular}
\end{table*}

We first include all previously spectroscopically identified galaxies at $z\sim 7.88$ in the A2744 field. These objects have previously been identified in NIRCam WFSS \citep{naidu_all_2024}, NIRSpec MSA \citep{morishita_early_2023,chen_jwst_2024,cameron_nebular_2024} and IFU \citep{hashimoto_reionization_2023, morishita_accelerated_2025}, and NIRISS WFSS \citep{roberts-borsani_early_2022}. These galaxies are included in the upper section of Table~\ref{tab:Targets}. Following this, we identify photometric candidates from the literature that have an SED consistent with being at $z\sim7.9$. These are taken from \cite{zheng_young_2014,morishita_early_2023,hashimoto_reionization_2023}. Finally, in producing the photometric catalogues we run the SED-fitting code \texttt{eazy} \citep{brammer_eazy_2008}, again following the steps described in \cite{weibel_galaxy_2024}. 

We note here that there is a clear spike at $z\sim 7.9$ in the photometric redshift distribution from \texttt{eazy}, which correlates with the protocluster. While the photometric redshift distribution of galaxies associated with overdense structures can often be diffuse due to the uncertainties in photometric redshift estimates, this tightly peaked redshift distribution ($z_{\rm FWHM}\sim 0.05$) underlines the revolutionary abilities of medium-band photometry in identifying and characterising overdense environments. 

We select objects with photometric redshifts lying around the peak in the photometric redshift distribution and consistent with the photometric redshifts and their associated uncertainties of our spectroscopically confirmed sample ($7.8 < z_{\rm phot} < 8.0$; corresponding to 60 cMpc). Following this, we remove SEDs that are dominated by noise. This often leads to the removal of galaxies at $m_{\rm F150W}\gtrsim 28.1$~AB. Our final sample consists of only high-confidence candidates with photometric redshift distributions with 16th and 84th percentiles that lie within the aforementioned redshift interval ($7.8 < z < 8.0$; $\sim 10\%$ of the full photometric sample). This criteria selects all of the previously spectroscopically-confirmed galaxies in \pc\ (see Table~\ref{tab:Targets}), as well as seven new photometric candidate PRGs, two of which are newly spectroscopically-confirmed in Appendix~\ref{app:spec}. Our final sample is detailed in Table~\ref{tab:Targets}. 

We note here that our selection produces a bias against observing UV-faint objects with weak emission lines that ultimately have a poorly constrained photometric redshift. However, given we wish to study the properties of galaxies that we are confident reside within the protocluster, we value the purity of our sample more than the completeness. At the redshift of the protocluster, the \OIII emission lines fall outside of the high transmission wavelength range of the F430M filter, but within the F444W filter, while the \Hb emission falls in the F430M filter. This produces a characteristic, roughly equal, F430M and F444W detection which strongly constrains the photometric redshift of our photometric candidates (see Table~\ref{tab:Targets}). While two of our candidates show very weak emission line fluxes (discussed further in Section~\ref{sec:(mini)-quenched}), their strong Lyman and Balmer breaks, and their very close proximity ($< 1$~pkpc) to other confirmed PRGs, allows us to confidently consider them to be at $z=7.88$. Moreover, two of our new candidate galaxies were recently observed, and spectroscopically confirmed, by the UNCOVER program (see Appendix~\ref{app:spec}).

\section{Photometric properties}
\label{sec:phot_props}

\begin{figure}
\centering
\includegraphics[width=0.5\textwidth]{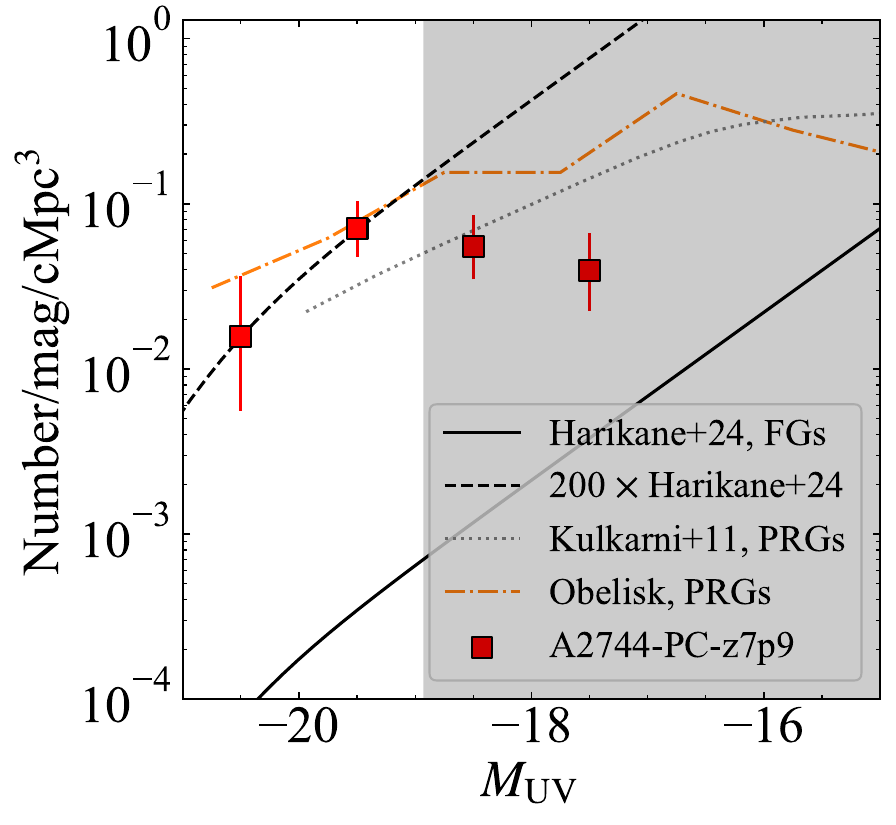}
\caption{The UV luminosity function (LF) of PRGs in our sample (red squares) compared to the ``nominal'' UVLF measured by \cite{harikane_pure_2024} for field galaxies (FGs; black solid line). We additionally compare to the UVLF from a semi-analytical model of an extremely overdense region at $z=8$ from \cite{kulkarni_reionization_2011} (grey dotted line) and the central $R=2$cMpc region of the Obelisk simulation at $z=7.9$ \citep{trebitsch_obelisk_2021} (orange dot-dash line). At the UV-bright end of the LF, the \pc\ region appears to be 200 times overdense relative to the nominal UVLF (black dashed line). There are indications of a mild turnover at $M_{\rm UV}>-18$, however our sample will become highly incomplete near the $5\sigma$ depth (indicated by the shaded grey region), as discussed in Section~\ref{sec:selection}, so this remains uncertain.}
\label{fig:LF}
\end{figure}

In the following section, we evaluate a number of galaxy properties that can be established without the need for SED-fitting: the UV luminosity, the UV slope and the Balmer break strength. The inferred UV luminosity function (LF) of galaxies resident in this protocluster is displayed in Figure~\ref{fig:LF}. This estimate utilises the magnitudes shown in Table~\ref{tab:Targets} and a volume defined as spanning a surface area of $21\ {\rm cMpc}^{2}$ (corresponding to the full spatial extent of protocluster-resident galaxies reported in Table~\ref{tab:Targets}) and a line-of-sight (LOS) distance of 6~cMpc (corresponding to $\Delta z\pm0.01$; the extent of the spectrosocpic redshift distribution). This volume is comparable to the expected sizes of protoclusters at $z\sim8$ \citep{chiang_galaxy_2017}. We also note that given the selection process, the LF is significantly impacted by a lack of completeness at $M_{\rm UV}>-19$ (indicated by the grey-shaded region). At magnitudes brighter than the completeness limit, the UVLF of \pc\ is compatible with the expectations of extremely overdense regions in semi-analytical models \citep{kulkarni_reionization_2011} and a protocluster environment from the cosmological simulation, Obelisk \citep[][]{trebitsch_obelisk_2021}\footnote{In Figure~\ref{fig:LF} we show the UVLF of the central $R=2$~cMpc Obelisk simulation at $z=7.9$, using the dust prescription, with an extinction law similar to the Small Magellanic Cloud, described in Trebitsch et al. in prep. and \cite{volonteri_exploring_2025}.}. An early plateau in the expected number of UV-faint galaxies is predicted in such overdense environments due to the enhanced UV-background suppressing star formation in low mass halos \citep[e.g.][]{zier_thesan-zoom_2025, katz_how_2020}, however, this is beyond the depths of the currently available imaging in this region. We do see a mild turnover in the UVLF occurring before the $5\sigma$ depth ($M_{\rm UV}\approx-18$), but note that the completeness function is not a step function and hence may be affecting the number count of galaxies already at $M_{\rm UV}<-18$. The \pc\ region appears to be at $\delta \sim 200$ relative to the nominal UVLF. However, it is important to note that in these most overdense environments there is a bias towards the most massive, UV-bright objects \citep{jespersen_significance_2025}, as is indicated by the variation in the slope of the UVLF between a protocluster core from the Obelisk simulations \citep{trebitsch_obelisk_2021} and that of field galaxies \citep{harikane_pure_2024}. As such, $\delta \sim 200$ likely corresponds to an upper bound on the overdensity of galaxies at lower UV magnitudes. The consistency between the observed UVLF and the Obelisk simulation at $z\sim7.9$ indicates that this region is consistent with the core of a region that will form a cluster of mass $M_{\rm vir}\sim 6.6 \times 10^{13} {\rm\ M_{\odot}}$ at $z=0$. When integrating the nominal UVLF down to the 5$\sigma$ imaging depth, using the volume defined above ($V \approx 125\ {\rm cMpc^3}$), we measure the overdensity of \pc\ to be $\delta \approx 70$, or $\delta \approx 60$ if we only consider spectroscopically-confirmed galaxies. These are consistent with the overdensity of \pc\ that was recently reported by \cite{morishita_metallicity_2025}. 

\begin{figure}
    \centering
    \includegraphics[width=0.99\linewidth]{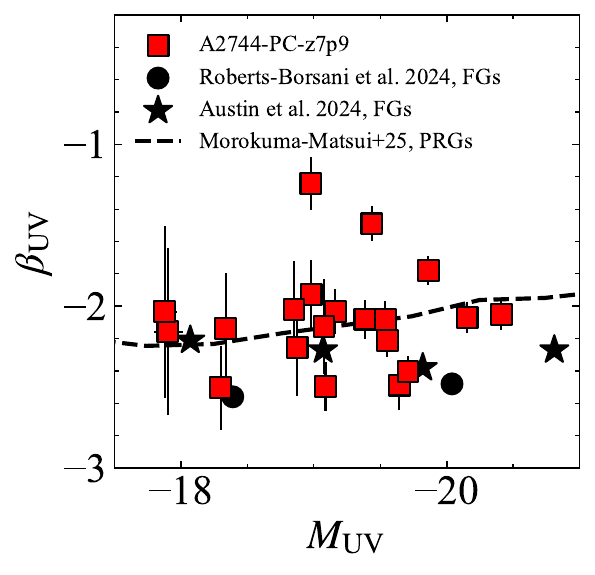}
    \caption{The UV continuum slopes of our sample (red squares) as a function of UV magnitude. We calculate the UV-slope by fitting the photometric fluxes in the filters covering the wavelength range of $0.13 \mu {\rm m}<\lambda_{\rm rest}<0.31 \mu {\rm m}$ with a power-law. We compare these to literature values of field galaxies (FGs) measured from photometry \citep[][black stars]{austin_epochs_2024} and stacked spectroscopy \citep[][black circles]{roberts-borsani_between_2024}. At the UV-bright end, our protocluster-resident sample lies above the $M_{\rm UV} - \beta$ relation seen in the field, and is largely inline with the relation seen in PRGs in simulations \citep[][black dashed line]{morokuma-matsui_forever22_2025}. This appears to indicate that our protocluster sample is more dust enriched than field galaxies in the literature, potentially evidencing their more evolved nature.}
    \label{fig:UV-slopes}
\end{figure}

The UV continuum slopes of our sample are evaluated by fitting the emission detected in the filters between $0.13 \mu {\rm m}<\lambda_{\rm rest}<0.31 \mu {\rm m}$ (F140M, F150W, F162M, F182M, F200W, F210M, F250M, F277W, F300M)  with a power law ($f_{\lambda}\propto \lambda^{\beta}$). We exclude the F115W filter as this is affected by neutral hydrogen in both the intergalactic medium (IGM) and on local scales producing significant damping around the Lyman-break. We additionally remove filters at longer wavelengths as the assumption of the continuum being governed by a power-law breaks down beyond the Balmer limit ($\lambda \approx 0.365 \mu$m). We utilise a Monte Carlo (MC) uncertainty propagation technique to estimate uncertainties in the inferred UV-slope, by repeatedly drawing the fluxes from a Gaussian centred on their nominal values and with a standard deviation given by their associated uncertainties. These values span a wide range from $-2.5 \leqslant \beta \leqslant -1.3$. The median UV-slope of our sample is $\beta = -2.1 \pm 0.3$, which is shallower than the value for the median stack in \cite{roberts-borsani_between_2024} of $\beta = -2.50 \pm 0.04$. The UV-slope as a function of UV-luminosity is displayed in Figure~\ref{fig:UV-slopes}, and the PRGs are preferentially redder than both the median values in spectroscopically observed galaxies \citep{roberts-borsani_between_2024} and photometrically observed sources \citep{austin_epochs_2024} in the field. While for UV-faint galaxies, the median UV-slope of our PRGs are consistent with the values for field galaxies in the literature, at the UV-bright end, the median UV-slope of PRGs differs by $\Delta\beta\approx 0.3$. We note that this difference between field galaxies and PRGs is beyond the difference in UV-slopes of observed galaxies in overdense and underdense environments ($\Delta\beta\approx 0.15$) at high redshift in \cite{li_epochs_2025}. This suggests that with increasing overdensity the effect on the UV-slope becomes more significant. Red UV-slopes are typically associated with either enhanced nebular emission \citep[e.g.][]{cameron_nebular_2024} or more dusty galaxies. Given that many galaxies in our sample show relatively weak emission line equivalent widths (see Figures~\ref{fig:SED+SFH1},~\ref{fig:SED+SFH2},~\ref{fig:SED+SFH3} and~\ref{fig:SED+SFH4}), increased nebular emission appears not to be a likely explanation. Combined with previous ALMA dust detections around some of our sample \citep[YD4, YD7-W and YD1,][the former two have the reddest UV-slopes in our sample]{laporte_dust_2017,hashimoto_reionization_2023}, the reddening of the UV-slope appears consistent with enhanced dust attenuation, which gives a first indication of the relatively evolved nature of our sample compared to typical galaxies at the same redshift. 

\begin{figure}
\centering
\includegraphics[width=0.5\textwidth]{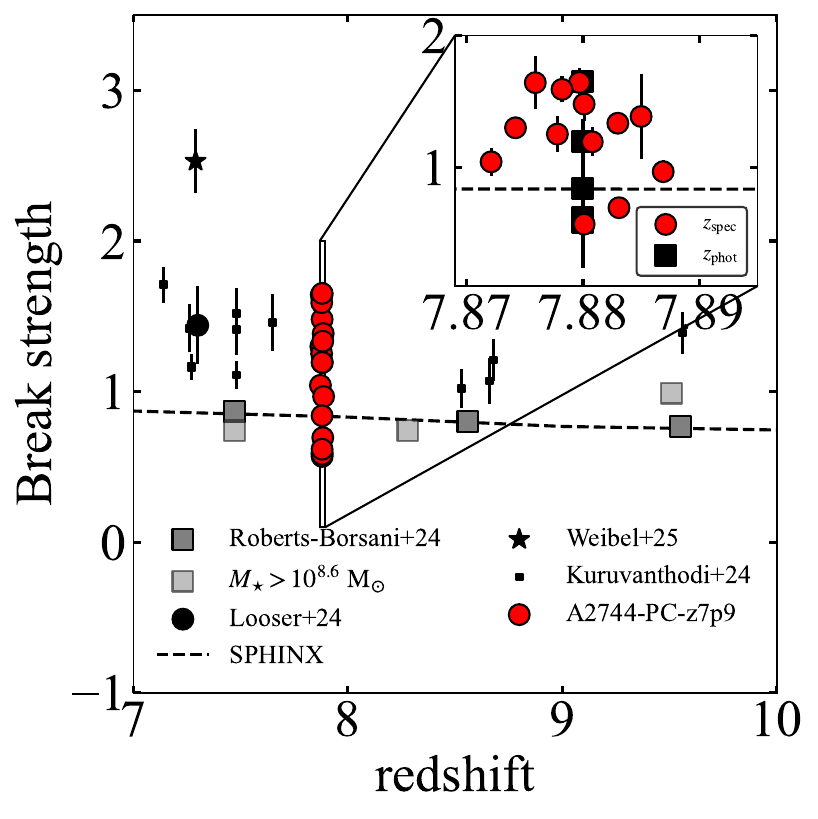}
\caption{The Balmer break strength measured as the ratio of $(f_{\rm \nu,460M}+f_{\rm \nu,480M})/(2\times f_{\rm \nu,277W})$ for our sample. In the main panel, the full sample is indicated by red circular points, while in the inset, the spectroscopic and photometric samples are differentiated by circles and squares, respectively. Note here that we assume a redshift of $z=7.88$ for the photometric data points. We include, for comparison, Balmer break measurements from galaxies that host strong Balmer breaks in their SEDs (black circle, \citealt{looser_recently_2024}; black star, \citealt{weibel_rubies_2025}; black squares, \citealt{kuruvanthodi_strong_2024}), and from stacked spectra, representing the general galaxy population (grey squares) and a subsample of galaxies with $M_{\star}> 10^{8.6}\ {\rm M_{\odot}}$ (transparent grey squares) from \citealt{roberts-borsani_between_2024}. We additionally show the Balmer break strengths seen in the \texttt{sphinx}$_{20}$ simulations \citep{rosdahl_sphinx_2018,rosdahl_lyc_2022,katz_sphinx_2023}, with the the black dashed line (taken from \citealt{witten_rising_2025}). The Balmer break strengths seen in many of the PRGs sit above those of typical galaxies at these redshifts, and are consistent with some of the strongest Balmer breaks strengths seen in the literature.}
\label{fig:BBs}
\end{figure}

The Balmer break is known to be a tracer of stellar ages \citep{bruzual_stellar_2003}. While there have been observations of strong Balmer breaks in compact, red galaxies (``Little Red Dots'') these are often associated with broad Balmer lines which additionally often show absorption features \citep{matthee_eiger_2023}. This leads to the conclusion that both the Balmer break and absorption of Balmer lines is likely associated with an extremely dense neutral hydrogen atmosphere surrounding a black hole \citep{ji_blackthunder_2025,deugenio_blackthunder_2025,naidu_black_2025}. However, these objects are distinguishable from star-forming galaxies thanks to their ``V''-shaped SEDs, which none of our sample show. Moreover, no previous studies of these galaxies have found any evidence for broad Balmer lines, meaning that the Balmer breaks present in our sample are likely of a stellar origin. 

The Balmer break strength is typically measured using spectroscopy by taking intervals just before and after the Balmer break ($\lambda \approx 0.365 \mu$m) that are free from emission line contamination. Estimating this with photometry is more challenging, as at wavelengths below the \OIII doublet, numerous emission lines are available to contaminate the continuum measurement (e.g. \OII, [\ion{Ne}{III}], H$\gamma$, H$\beta$). However, in the wavelength range shortly following the \OIII doublet there exist no emission lines capable of significantly contaminating photometric continuum measurements. At $z=7.88$ the \OIII doublet falls within the F430M and F444W filters, leaving the average of F460M and F480M filters available for rest-optical continuum measurements. The measurement of the continuum below the Balmer break is simpler given the lack of emission lines in the wavelength range directly below the break and as such we take the F277W filter as the continuum level pre-Balmer-break. The inferred Balmer break strengths ($B$) of our sample are shown in Figure~\ref{fig:BBs}. While two of the galaxies in our sample, both located outside of the most clustered regions, show definitive Balmer jumps (inverse Balmer breaks; $B<1$) -- evidencing their young, highly star-forming nature, driving strong nebular emission -- the majority of our sample show considerably stronger Balmer breaks than is typical in field galaxies at this redshift in both stacked spectra \citep{roberts-borsani_between_2024} and simulations \citep{rosdahl_sphinx_2018,rosdahl_lyc_2022,katz_sphinx_2023}. Moreover, the break strengths of our sample are stronger than those observed in stacks of high-mass ($M_{\star}> 10^{8.6}\ {\rm M_{\odot}}$) galaxies from \cite{roberts-borsani_between_2024}, suggesting that these strong break strengths are abnormal even in high stellar mass galaxies. The sample shows Balmer break strengths that are consistent with large breaks identified in the literature at $z\sim7-10$ \citep{looser_recently_2024,kuruvanthodi_strong_2024}. 

\section{SED-Fitting}
\label{sec:sed-fitting}

\begin{figure}
    \centering
    \includegraphics[width=0.99\linewidth]{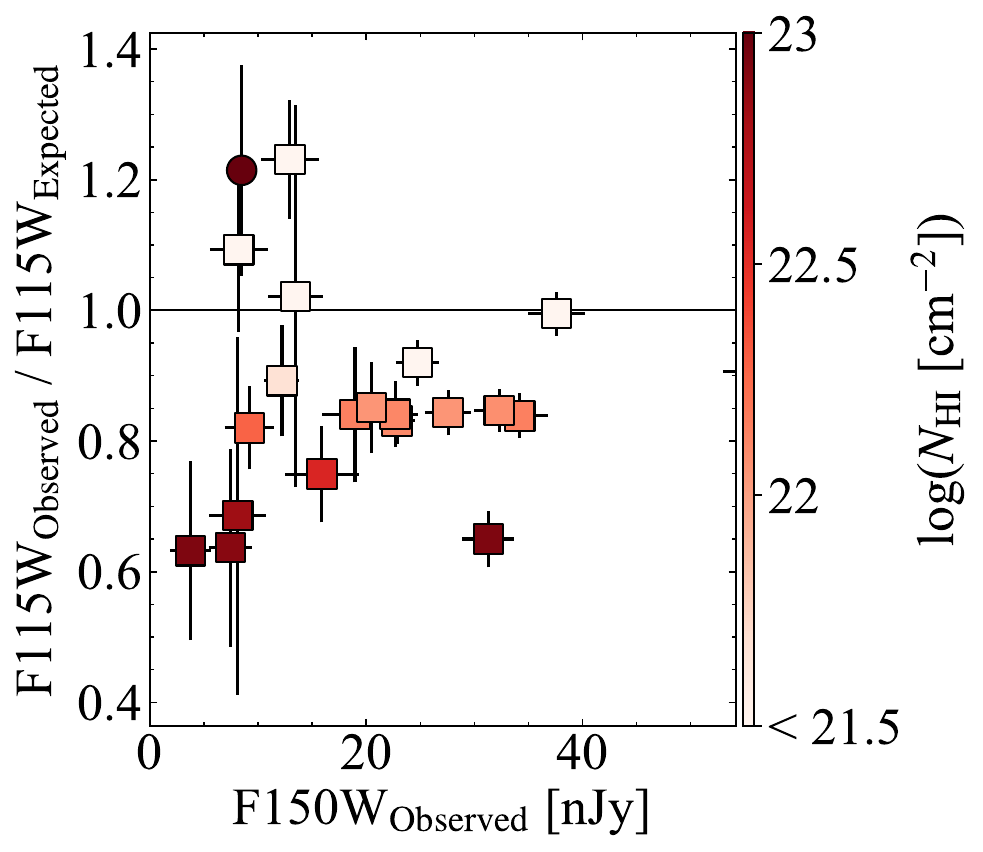}
    \caption{The ratio of the flux observed in the F115W filter to the expected F115W from our \texttt{PROSPECTOR} SED-fitting, described in Section~\ref{sec:sed-fitting}. The data points are colour coded by the inferred neutral hydrogen column density, discussed in Section~\ref{sec:sed-fitting}. We note here the one circular data point is simultaneously a strong LAE and DLA, and hence we utilise the spectroscopically measured $N_{\rm HI}$ from Witten et al. (in prep.). The majority of galaxies show a strong suppression in the F115W filter that is best explained by high neutral hydrogen column densities ($N_{\rm HI}>10^{22}\ [{\rm cm^{-2}}]$), which has been seen in the galaxies for which deep NIRSpec prism observations are available \citep{chen_jwst_2024, witten_rising_2025, mason_constraints_2025}.}
    \label{fig:F115W}
\end{figure}

\begin{table*}
\centering
\caption{The results of SED-fitting the PRGs, where SFR$_{10}$ and SFR$_{100}$ denote the average star-formation rate (SFR) in the previous 10 and 100~Myr. Note that, as with all galaxy properties reported in this paper, these values are corrected for magnification. Galaxies have been separated into core and non-core-resident galaxies, as defined in Section~\ref{sec:proto_props}.}
\label{tab:Properties}
 \begin{tabular}{lcccc}
  \hline \\[-0.3 cm]
  Name & $\log_{10} (N_{\rm HI}\ {\rm [cm^{-2}]})$ & log($M_{\star}\ [{\rm M_{\odot}}]$) & SFR$_{10}\ [{\rm M_{\odot}/yr}]$ & SFR$_{100}\ [{\rm M_{\odot}/yr}]$ \\[0.1 cm]
  \hline
  \multicolumn{3}{l}{\bf Core-resident galaxies}&\\
  YD7-E & $ 22.1 \pm 0.3 $ & $9.17^{+0.06}_{-0.06}$& $1.68^{+1.65}_{-1.10}$ & $4.87^{+4.98}_{-2.69}$ \\[0.1 cm]
  ZD6 & $22.1 \pm 0.5$ & $9.08^{+0.04}_{-0.06}$ & $2.35^{+1.28}_{-0.93}$ & $2.24^{+2.46}_{-1.25}$ \\[0.1 cm]
  YD4 & $22.9 \pm 0.1$ & $9.04^{+0.17}_{-0.17}$ & $11.82^{+4.85}_{-5.38}$ & $4.75^{+5.55}_{-2.75}$ \\[0.1 cm]
  ZD12-E & $<21.5$ & $8.96^{+0.05}_{-0.05}$ & $1.03^{+1.88}_{-0.88}$ & $3.33^{+4.42}_{-1.95}$ \\[0.1 cm]
  YD6 & $22.1 \pm 0.8$ & $8.84^{+0.10}_{-0.11}$ & $4.85^{+4.15}_{-2.46}$ & $3.19^{+4.17}_{-1.99}$ \\[0.1 cm]
  YD1 & $22.2 \pm 1.0$ & $8.74^{+0.19}_{-0.19}$ & $9.12^{+4.29}_{-3.13}$ & $2.02^{+1.83}_{-1.05}$  \\[0.1 cm]
  YD7-W & $22.3 \pm 0.5$ & $8.65^{+0.14}_{-0.25}$ & $8.61^{+9.42}_{-4.19}$ & $2.28^{+3.07}_{-1.72}$  \\[0.1 cm]
  ZD3 & $22.0 \pm 0.3$ & $8.62^{+0.16}_{-0.09}$ & $2.84^{+3.21}_{-1.42}$ & $1.81^{+3.75}_{-1.28}$  \\[0.1 cm]
  ZD12-W & $22.6 \pm 0.3$ & $8.55^{+0.10}_{-0.13}$ & $2.40^{+0.96}_{-0.70}$ & $0.99^{+0.60}_{-0.50}$ \\[0.1 cm]
  ZD1 & $22.9  \pm 0.6 $ & $8.24^{+0.09}_{-0.11}$& $0.41^{+0.38}_{-0.27}$ & $0.59^{+0.59}_{-0.33}$ \\[0.1 cm]
  s1 & $22.8 \pm 1.2$ & $7.35^{+0.17}_{-0.19}$ & $1.04^{+0.41}_{-0.28}$ & $0.13^{+0.10}_{-0.05}$  \\[0.1 cm]
  \hline
  \multicolumn{3}{l}{\bf Non-core-resident galaxies}&\\
  ALT-41799 & $22.2 \pm 0.3$ & $9.08^{+0.09}_{-0.07}$& $10.70^{+4.55}_{-3.46}$ & $3.43^{+2.68}_{-1.69}$ \\[0.1 cm]
  PC2-W  & $ 22.1 \pm 0.3 $ & $8.39^{+0.09}_{-0.07}$& $2.33^{+1.03}_{-0.81}$ & $1.54^{+0.84}_{-0.65}$ \\[0.1 cm]
  ZD4 & $23.0\pm0.3 \ ^a$ & $8.27^{+0.14}_{-0.28}$ & $1.66^{+1.01}_{-0.83}$ & $0.62^{+0.74}_{-0.41}$  \\[0.1 cm]
  ZD2 & $<21.5$  & $8.21^{+0.22}_{-0.04}$ & $14.44^{+1.62}_{-0.71}$ & $1.45^{+0.58}_{-0.08}$ \\[0.1 cm]
  YD8 & $<21.5$ & $8.17^{+0.22}_{-0.13}$ & $4.88^{+1.96}_{-1.19}$ & $0.92^{+0.89}_{-0.41}$  \\[0.1 cm]
  z8-2 & $<21.5$ & $8.12^{+0.14}_{-0.07}$ & $11.10^{+4.76}_{-1.07}$ & $1.3^{+0.6}_{-0.2}$  \\[0.1 cm]
  PC3  & $ <21.5 $ & $8.07^{+0.15}_{-0.10}$& $7.70^{+1.63}_{-2.28}$ & $0.91^{+0.62}_{-0.35}$  \\[0.1 cm]
  PC2-E & $ 21.7 \pm 1.0 $& $7.93^{+0.13}_{-0.14}$& $2.78^{+0.67}_{-0.60}$ & $0.38^{+0.20}_{-0.12}$  \\[0.1 cm]
  PC4  & $ <21.5 $ & $7.83^{+0.16}_{-0.20}$& $2.05^{+0.98}_{-0.81}$ & $0.36^{+0.35}_{-0.21}$  \\[0.1 cm]
  PC1 & $ <21.5 $ & $7.53^{+0.23}_{-0.26}$& $1.26^{+0.54}_{-0.32}$ & $0.18^{+0.19}_{-0.08}$  \\[0.1 cm]
  PC5 & $ 22.9 \pm 0.5 $ & $7.20^{+0.15}_{-0.18}$& $0.95^{+0.37}_{-0.18}$ & $0.10^{+0.09}_{-0.03}$  \\[0.1 cm]
  PC6 & $<21.5$ & $7.18^{+0.39}_{-0.16}$ & $0.94^{+0.71}_{-0.37}$ & $0.10^{+0.11}_{-0.04}$  \\[0.1 cm]
  \hline
 \end{tabular}
\vspace{0.5em}

\begin{minipage}{\textwidth}
$^a$\small\ The $N_{\rm HI}$ of ZD4 has been calculated from its spectrum as it shows strong Lyman-$\alpha$ (Ly$\alpha$) emission that contaminates the F115W filter (Witten et al. in prep.).
\end{minipage}
\end{table*}

In order to estimate the physical properties of the galaxies within the protocluster, we use the SED-fitting code \texttt{PROSPECTOR} \citep{johnson_prospector_2019,johnson_stellar_2021}. We employ the same 16-parameter model, including a flexible dust attenuation law \citep{conroy_propagation_2009} and a non-parametric SFH with a continuity prior \citep{leja_how_2019}, as utilised in \cite{witten_rising_2025}. 

We fit the aforementioned customised aperture photometry of our sample, utilising all but one of the available filters, the F115W filter. This exclusion is made to counteract the impact of heavy damping that is possible with extreme densities of neutral hydrogen on local scales (damped Lyman-$\alpha$ absorption, DLA; e.g. \citealt{heintz_strong_2024}), which is not modelled in \texttt{PROSPECTOR}. This removes a constraint on the photometric redshift for our photometric candidate sample. However, given that we strongly suspect these objects to be resident in the protocluster, in part thanks to their strong emission line detections in medium band filters, we simply fix the redshift of our \texttt{PROSPECTOR} fits to be at $z = 7.88$. Excluding the F115W filter ultimately improves the overall fit of our \texttt{PROSPECTOR} models, as SED-fitting codes often assume that the excessive suppression of flux in the filter covering the Lyman-break is caused by significant dust attenuation, thus producing a sub-optimal fit to the UV-continuum. 

The key results of the SED-fitting of each galaxy are reported in Table~\ref{tab:Properties}. The inferred SFHs of our sample vary from objects that are best-fit by instantaneous bursts through to extended SFHs over more than 100~Myr. This range of SFHs can be seen in the comparison of the SFR$_{10}$ and SFR$_{100}$ reported in Table~\ref{tab:Properties} -- these range from SFHs dominated by a recent ($<10$ Myr) burst (SFR$_{10}>>\ $SFR$_{100}$) and those recently declining or plateauing (SFR$_{10}\lesssim\ $SFR$_{100}$) (also see Figures~\ref{fig:SED+SFH1},~\ref{fig:SED+SFH2},~\ref{fig:SED+SFH3} and~\ref{fig:SED+SFH4}). Those galaxies that have declining or plateauing SFHs tend to be the most massive (log($M_{\star}[{\rm M_{\odot}}])\sim 9$), while those at the low mass end are exclusively characterised by extreme, recent bursts of star formation. 

It should be noted that SED-modelling with non-parametric SFHs is only capable of constraining the most recent burst in the SFH of galaxies. This is due to the ``outshining'' effect \citep{papovich_stellar_2001, pforr_recovering_2012, conroy_modeling_2013, tacchella_jwst_2023, whitler_star_2023,witten_rising_2025, wang_population_2025}, where young stellar populations are significantly brighter than an equal mass older stellar population. However, this most recent burst can be combined with other constraints on the maturity of the galaxy, such as the presence of dust or chemical enrichment, in order to understand whether this burst event is indeed the only significant burst event in the galaxy's SFH. 

The outshining effect generally produces galaxies with Balmer jumps at the highest redshifts \citep[e.g.][]{roberts-borsani_between_2024}. The presence of a Balmer break is therefore both rare, and necessitates significant star formation for extended periods ($\sim 100$~Myr) and little ongoing star formation \citep{looser_recently_2024, witten_rising_2025}. As such, the Balmer break strengths seen in our sample require that many of the PRGs host significant old stellar populations, consistent with the extended SFHs inferred from our SED-fitting (see Figure~\ref{fig:SFHs}). 

\subsection{Neutral hydrogen column densities}
\label{ssec:NHi}

The excessive damping of the Lyman-break caused by extremely dense neutral hydrogen on local scales has previously been noted for galaxies within this protocluster \citep{chen_jwst_2024, mason_constraints_2025}. This damping is so significant that it causes a suppression of flux in wide band filters around the Lyman-break, far beyond the IGM attenuation curve \citep{madau_radiative_1995}, and ultimately led to the overestimation of the redshift of galaxies in the protocluster \citep[e.g.][]{zheng_young_2014, laporte_dust_2017}. This damping can be seen in the discrepancy between the F115W flux of our best-fit \texttt{PROSPECTOR} model (we remind the reader that the F115W filter is removed from the fitting procedure for this exact reason) and the observed F115W flux, shown in Figure~\ref{fig:F115W}. 

When the density of the neutral gas is sufficiently high, as in DLAs, the optical depth is a function only of $N_{\rm HI}$. As such, by utilising an approximation of the Voigt-Hjerting function, presented in Equation 9 of \cite{smith_lyman_2015}, following \cite{heintz_strong_2024}, one can estimate the neutral hydrogen column density of the DLA. We apply this transmission function, combined with the IGM transmission\footnote{We estimate the IGM transmission following \cite{mason_measuring_2020} assuming no ionised bubble and a neutral fraction of one, hence maximising the effect of IGM attenuation. Our measurement of $N_{\rm HI}$ is therefore a lower bound.}, to the \texttt{PROSPECTOR} model spectrum at varying $N_{\rm HI}$ until the observed F115W flux is recovered. We use MC uncertainty propagation to find the optimal $N_{\rm HI}$ for a Gaussian distribution around the observed F115W flux and its uncertainties. This model is a simplification of the physical scenario as \Lya emission can be present \citep[as discussed in][]{heintz_jwst-primal_2025}. However, given the assumption that no \Lya is present, the inferred $N_{\rm HI}$ represents a lower bound on the column density of neutral hydrogen in the DLA. In reality, only one of the spectroscopically observed galaxies in A7244-PC, ZD4, is known to show strong \Lya emission \citep{chen_jwst_2024,cameron_nebular_2024}, and this is one of the few objects in Figure~\ref{fig:F115W} that is inconsistent with having an observed flux at or below the expected flux. Given this, in Figure~\ref{fig:F115W} we show the spectroscopic $N_{\rm HI}$ measurement for ZD4 (Witten et al. in prep.). Only when $N_{\rm HI}\gtrsim 10^{21.5}\ {\rm cm^{-2}}$ does a notable variation in the F115W flux occur ($\sim 10\%$), and as such we only report the column density of objects above this threshold. These measured column densities are reported in Table~\ref{tab:Properties}. 

\subsection{Star-forming main sequence}

\begin{figure}
\centering
\includegraphics[width=0.5\textwidth]{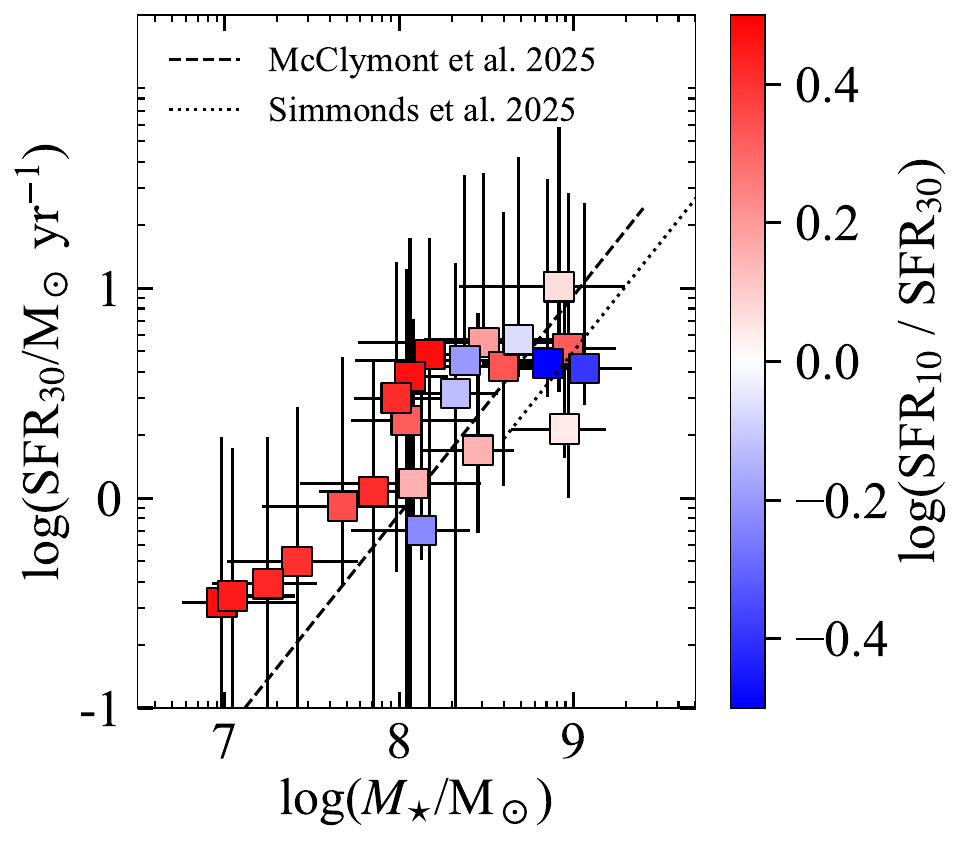}
\caption{The star-forming main sequence (SFMS) of our PRGs relative to the SFMS, derived from simulations \citep[][black dashed lines]{mcclymont_thesan-zoom_2025} and from observations from a large population study \citep[][black dotted line]{simmonds_bursting_2025}, of field galaxies. The average SFR over the previous 30~Myr is chosen as this is informed by both emission line strengths and UV continuum strength, while the stellar mass is largely informed by the optical continuum. In addition, the colour-coding of the squares shows the ratio of the very recent (SFR$_{10}$) to slightly longer term (SFR$_{30}$) star formation rates. While the lower-mass galaxies are typically elevated relative to the SFMS, due to selection effects, we see little evidence of highly bursting, high-mass galaxies, and instead the most massive galaxies appear to be mildly below the SFMS from simulations, with declining SFHs over the last 30 Myr.}
\label{fig:SFMS}
\end{figure}

The star-forming main sequence (SFMS) characterises the growth of galaxies, highlighting those that either have enhanced or suppressed star formation relative to the average population at a given stellar mass and redshift. In protocluster environments it is expected that the higher than average UV-background suppresses star formation in low-mass satellite halos \citep[e.g.][]{katz_how_2020,borrow_thesan-hr_2023,zier_thesan-zoom_2025}. In Figure~\ref{fig:SFMS} we show the SFMS for our sample with the average SFR over the previous 30~Myr, as diagnosed by our \texttt{PROSPECTOR} SED-fitting in Section~\ref{sec:sed-fitting}. We chose this intermediate timescale, relative to the typical 10 or 100~Myr average, as the star formation over the last $\sim 30$ Myr largely defines the UV and emission line strengths in galaxies \citep[e.g.][]{mcclymont_thesan-zoom_2025}. Instead, features like the Balmer break give information on the star formation over longer timescales ($\sim 100$~Myr), which at this redshift informs the total stellar mass. We compare in Figure~\ref{fig:SFMS} to the SFMS of field galaxies from the \textsc{thesan-zoom} simulations \citep{mcclymont_thesan-zoom_2025, kannan_introducing_2025} and from \texttt{PROSPECTOR} fitting of a large sample of galaxies from \cite{simmonds_bursting_2025}. We compare this protocluster environment to the SFMS of field galaxies in order to highlight any variations from the typical SFMS. The PRGs largely follow the slope of the SFMS of field galaxies from both simulations and observations, which is expected from the analysis of PRGs in simulations \citep{morokuma-matsui_forever22_2025}. The results of \cite{simmonds_bursting_2025} suggest their sample suffers from incompleteness below $M_{\star}\approx 10^{8.6}\ \rm{M_{\odot}}$ due to the enhanced burstiness of high-redshift galaxies. However, at the high-mass end where we are complete, we find no objects that lie significantly above the SFMS found in \textsc{thesan-zoom}. Instead many of these galaxies fall below the SFMS, moreover, many have plateauing or declining SFRs over the last 10 Myr relative to the last 30 Myr. While this may appear like a suppression of star formation exclusively in the most massive galaxies, some lower mass objects also show recently declining SFHs, as such, it is important to consider what may be driving these declining star-formation rates. 

\subsection{(Mini-)quenched galaxies}
\label{sec:(mini)-quenched}

\begin{figure}
\centering
\includegraphics[width=0.49\textwidth]{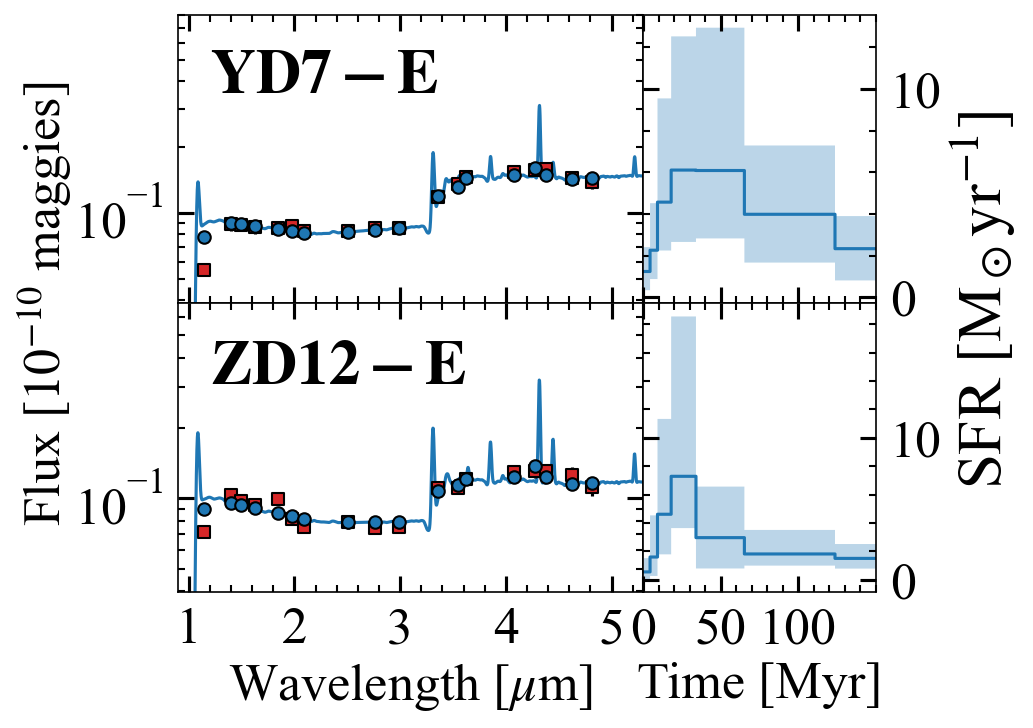}
\caption{Two candidate (mini-)quenched galaxies in the protocluster, exhibiting strong Balmer breaks and very weak emission lines. Their model spectra and photometry (blue line and circles) and observed photometry (red squares) are shown in the left-hand panels, while their best-fit SFH (blue line) and associated uncertainty (shaded blue region) from \texttt{PROSPECTOR} are shown in the right-hand panels. The SFHs of these objects are best characterised by a sustained period of star formation, over more than 100 Myr, followed by a recent decline, over the last 10 Myr, that has reached SFR$~\sim0\ {\rm M_{\odot}\ yr^{-1}}$ by the time of observation. Both objects reside within the highly clustered regions of the protocluster providing a possible hint at the suppression of star formation within the innermost region.}
\label{fig:(mini)-quenched}
\end{figure}

\begin{figure*}
\centering
\includegraphics[width=1\textwidth]{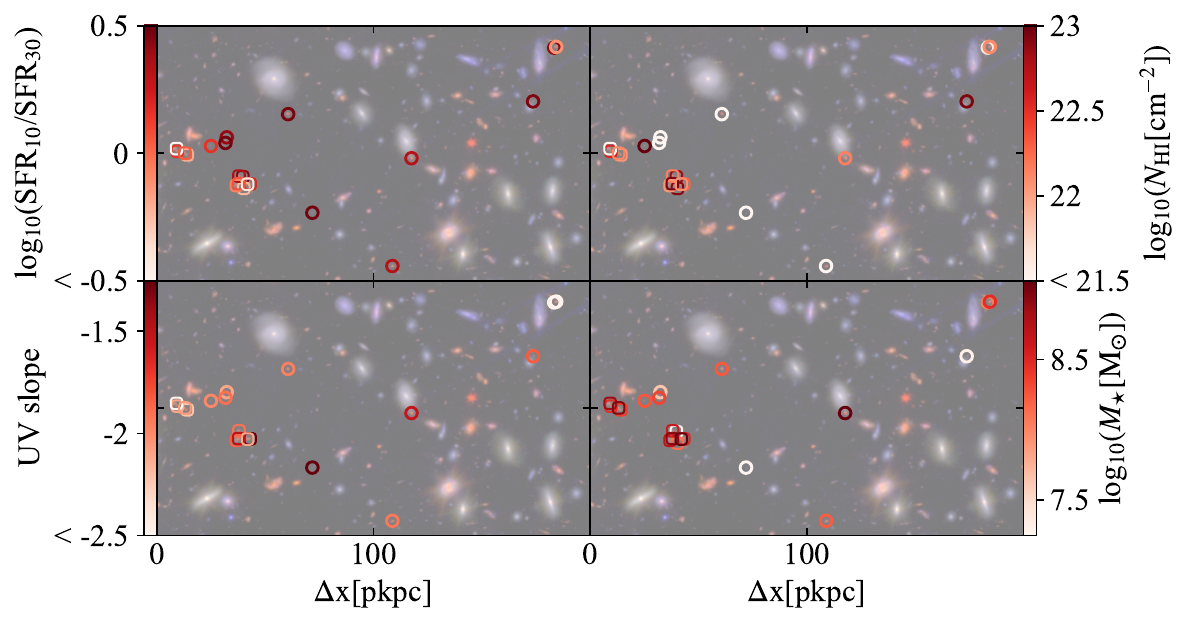}
\caption{A mapping of the properties of PRGs onto the RGB image of the protocluster, as seen in Figure~\ref{fig:RGB}. Each galaxy is shown by an unfilled point, where the colour indicates the galaxy property that is studied in each panel. Circular points indicate ``non-core'' galaxies, while square points indicate ``core'' galaxies. Each panel includes a zoom of the most clustered region of the protocluster, for ease of viewing. The following galaxy properties are shown: (top left:) The logarithm of the ratio of the recent (average over the last 10~Myr) to slightly more extended (average over the last 30~Myr) SFRs; (top right:) The logarithm of the neutral hydrogen column densities; (bottom left:) the slope of the UV continuum, $\beta$; (bottom right:) The logarithm of the stellar mass. The mapping of these galaxy properties onto the protocluster, indicates that the two most clustered regions tend to host the most massive, dusty galaxies that have declining star-formation histories and extreme neutral hydrogen column densities. The more sparsely distributed galaxies tend to be less massive, highly star-forming objects.}
\label{fig:map}
\end{figure*}

With this in mind, we note here the detection of two relatively massive ($M_{\star}\sim 10^{9} \rm M_{\odot}$) galaxies, YD7-E and ZD12-E, that host significant Balmer breaks ($B = 1.65\pm0.08$ and $B = 1.59\pm0.10$, respectively) and very weak nebular emission lines, shown in Figure~\ref{fig:(mini)-quenched}. These weak emission lines have previously been observed by \cite{venturi_gas-phase_2024} for YD7-E and by \cite{morishita_accelerated_2025} for ZD12-E (the measured equivalent width is comparable to that of our best-fit SED model in Section~\ref{sec:sed-fitting}).

YD7-E and ZD12-E appear consistent with the (mini)-quenched galaxies identified by \cite{looser_recently_2024,looser_jades_2025,trussler_like_2025,covelo-paz_systematic_2025,baker_zapped_2025}. These objects have stellar masses in the regime where permanent quenching is unlikely, however, they also lie above the typical mini-quenching mass regime. \cite{gelli_quiescent_2023} showed that galaxies in the SERRA simulations \citep{pallottini_survey_2022} with $M_{\star}>10^{9}\ {\rm M_{\odot}}$ spend 99\% of their time in an active phase. However, \citet{mcclymont_thesan-zoom_2025} show that even massive galaxies in the \textsc{thesan-zoom} simulations are bursty at high redshift due to rapid gas inflow from the IGM. This picture is consistent with the galaxies considered in this work, considering that the protocluster environment and high column density filaments are likely to facilitate such rapid inflows. Indeed, recent work by \cite{covelo-paz_systematic_2025} suggests that (mini)-quenched galaxies span a wider mass distribution than originally thought and hence we conclude that these objects are most likely to only be temporarily quenched. 

These two objects are resident in extremely clustered regions within the protocluster. In such a clustered environment, interactions between galaxies can either significantly enhance or suppress star formation {\it via} gravitational or feedback mechanisms \citep[e.g.][]{witten_deciphering_2024}. While such merger activity is not expected to be extremely frequent for field galaxies \citep{puskas_constraining_2025, duan_galaxy_2025}, within the core of a protocluster, the density of objects means that this will not only be more frequent \citep{marcelin_enhanced_2025}, but these interactions may affect many objects simultaneously, producing synchronised SFHs. The identification of galaxies with declining SFHs within the protocluster core motivates a broader investigation of the variation in the properties of galaxies as a function of position within the protocluster.

\section{Protocluster properties}
\label{sec:proto_props}

Simulations suggest that at $z>5$ protocluster environments should undergo an inside-out growth phase, with the majority of star-formation localised in a dense core of a few 10's pkpc \citep{chiang_galaxy_2017,lim_flamingo_2024}. Within that context, it is important to study the variation in galaxy properties as a function of spatial position within the protocluster. In order to do so, it is key to identify the most clustered region of the protocluster -- the protocluster core. While identifying the hot halo is feasible at lower redshifts with X-rays or the Sunyaev–Zel’dovich effect, at high redshift this measurement is not currently possible, and hence we must utilise alternative probes. The core should be one of the most evolved regions of the protocluster, forming its first stars early, and with significant gas inflows leading to accelerated galaxy evolution. The detection of dust continuum \citep{laporte_dust_2017, hashimoto_reionization_2023} and evidence for an old stellar population within YD4 \citep{witten_rising_2025}, marks the region around YD4 including YD7-E/W, YD6, YD1, s1 and ZD1 as a strong candidate for the core of this protocluster. A second highly clustered region was recently studied by \cite{morishita_accelerated_2025}, appearing to host numerous clumps that have a wide range in gas-phase metallicities; this region includes ZD12-E/W, ZD3 and ZD6. These two highly clustered regions are separated by just $\sim 20$ pkpc and hence constitute our ``core'' galaxies \citep[this spatial extent is in line with predictions for protocluster core sizes from][]{chiang_galaxy_2017, lim_flamingo_2024}. The remaining galaxies studied in this work are largely seen to be more sparsely distributed on the sky, albeit many have a close companion galaxy. 

With this spatial distribution in mind, we overplot the galaxy properties on an RGB image of the protocluster in Figure~\ref{fig:map}. This figure hints at a distinction between the properties of galaxies within these clustered regions (core galaxies) and the more sparsely distributed population (non-core galaxies). The core galaxies are mostly characterised by having large stellar masses and plateauing or declining star formation histories. Moreover, they host large neutral hydrogen column densities, and for some of the galaxies within the region around YD4, relatively red UV-slopes. These observations point towards these galaxies residing within regions of the cosmic web that are subject to significant gas accretion and early, sustained star formation that results in evolved stellar populations, with galaxies showing red slopes and strong Balmer breaks that diagnose their dusty, older stellar populations. Non-core galaxies can largely be characterised as having SFHs that are rapidly increasing, suggesting these galaxies are relatively young and, at least currently, are undergoing a large burst of star formation. The stellar masses and neutral hydrogen column densities of these objects vary significantly. Simulations \citep[e.g.][]{Boylan-Kolchin_resolving_2009, trebitsch_obelisk_2021, bennett_growth_2024} and lower redshift observations \citep[e.g.][]{castignani_virgo_2022} suggest that the neutral gas should be distributed in a filamentary structure with high column densities within the core of the protocluster and varying densities in the outskirts. Our observations appear consistent with this picture, whereby this pristine neutral gas is not distributed equally on large-scales.

\begin{figure}
\centering
\includegraphics[width=0.5\textwidth]{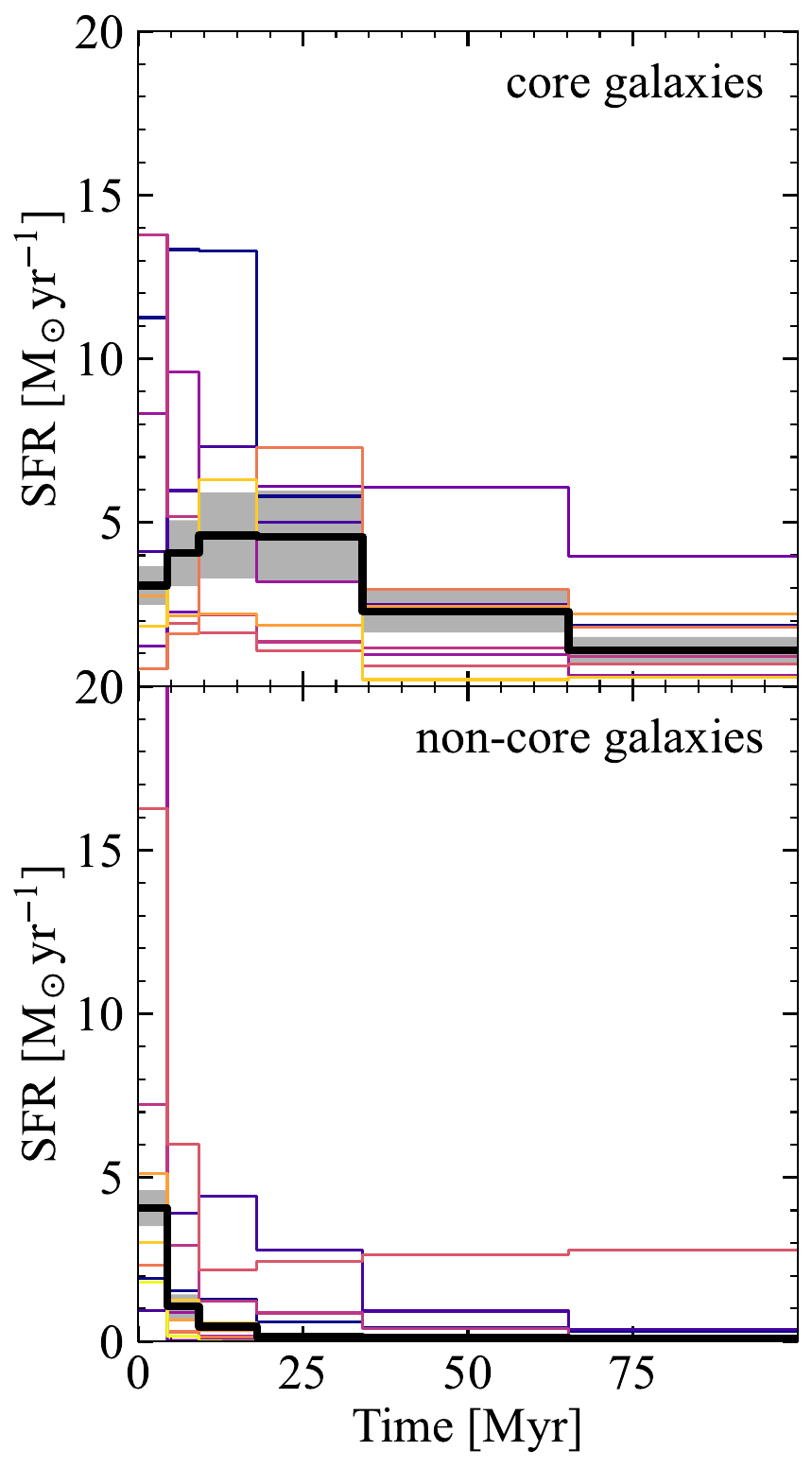}
\caption{The star-formation histories of galaxies resident within the most clustered regions of the protocluster, ``core'' galaxies (top panel), compared to those in the more sparsely populated regions, ``non-core'' galaxies (bottom panel). The SFHs of each galaxy are indicated by a coloured line. We indicate the median star-formation rate in each time bin with the black line and the associated MC error shaded in grey. The ``core'' galaxies exhibit more extended SFHs, with a declining SFR over the previous $\sim 10$ Myr, while ``non-core'' galaxies are characterised by strong ongoing bursts that largely began over the last $\sim 20$ Myr.}
\label{fig:SFHs}
\end{figure}

To place this difference between core and non-core galaxies in perspective, in Figure~\ref{fig:SFHs} we show the individual and median stacked SFHs of these populations. The core galaxies clearly evidence a more extended SFH, requiring star formation to commence more than 100~Myr before the time of observation. The median SFH of these objects also exhibits a turnover in its SFR, plateauing between $10-30$~Myr in the past and with a declining SFR in the final 10~Myr. This synchronisation in the SFHs of core galaxies likely evidences the synchronisation of gas kinematics within these densely clustered regions (see Fudamoto et al. in prep. for further discussion). The non-core galaxies are dominated by significant, ongoing bursts of star formation, that have largely only commenced in the last $< 15$~Myr. There are likely significant selection effects at play, as these objects largely have lower stellar masses, and hence we are biased to observe only the most star-forming of these objects. However, at high masses, where we expect to be relatively complete, the weakly star-forming objects are confined to the core regions. This is further highlighted by the fraction of the total star formation occurring within the core of the protocluster, which has evolved from dominating the total star formation rate of the protocluster over the preceding 100~Myr ($\sim 70\%$) to being subdominant in the most recent 10~Myr ($\sim 40\%$).

The identification of two (mini)-quenched galaxies and a recent decline in the star-formation history of core galaxies are likely linked to the lack of evidence for active galactic nuclei (AGN) activity within the core (discussed in Section~\ref{sec:phot_props}). The processes driving the reduction in the star-formation rate of these objects is potentially tied to the lack of AGN activity detected, despite the large stellar masses of many of these objects. Instead, either stellar or AGN feedback, or indeed gravitational interactions, have helped to expel gas from the interstellar medium, halting star formation and significant accretion onto any supermassive black hole. Notwithstanding the likely recent feedback episodes, a large neutral hydrogen reservoir persists, as evidenced by the high neutral hydrogen column densities of core-resident galaxies.

Our characterisation of the distribution of galaxy properties as a function of position within the protocluster supports a picture where core-resident galaxies are at a more evolved stage, and even have declining star-formation histories, while non-core galaxies have highly bursting SFHs. This is at odds with the early inside-out growth phase, presented by \cite{chiang_galaxy_2017}. These properties instead suggest that the protocluster is already at a relatively advanced stage of its evolution, with a mature core, that is slowing in its rate of star-formation, with much more rapidly star-forming galaxies found outside of the core, which is expected to be in place at later cosmic epochs ($z\lesssim 5-6$).

Moreover, the total stellar mass enclosed within the $R\sim 100$ pKpc region ($M_{\star}\sim 10^{10}\ \rm{M_{\odot}}$), is considerably higher than the values expected from the FLAMINGO simulations \citep{kugel_flamingo_2023,schaye_flamingo_2023} within the same radius (${\rm log_{10}}(M_{\star}[\rm{M_{\odot}}])= 8-9$; \citealt{lim_flamingo_2024}) and in line with the total stellar mass expected from a much larger volume ($R\sim7-14$ cMpc) from the FOREVER22 simulations \citep{yajima_forever22_2021,morokuma-matsui_forever22_2025}. \cite{lim_flamingo_2024} note that the total stellar mass of a high-redshift protocluster is independent of the ultimate cluster mass at $z=0$. Instead, the total stellar mass within $R\sim100$ pKpc is consistent with their results at $z=6$ for protoclusters that have halo masses of $M_{\rm halo}\sim 10^{12}\ {\rm M_{\odot}}$. These results suggest that \pc\ is relatively evolved, resident within a halo of mass $M_{\rm halo}\sim 10^{12}\ {\rm M_{\odot}}$, consistent with the recently updated halo mass estimate from \cite{morishita_metallicity_2025}. Utilising the adapted stellar-to-halo-mass relation from \cite{witten_not_2025}, and the total stellar mass within the most massive $R\sim 30\ {\rm pkpc}$ area, we estimate the halo mass of \pc\ to be ${\rm log_{10}}(M_{\rm halo}[\rm M_{\odot}]) = 11.6 \pm 0.2$.

\section{Conclusions}
\label{sec:conclusions}

Utilising recent deep and relatively high spectral resolution NIRCam imaging of the Abell 2744 lensing field \citep{suess_medium_2024,bezanson_jwst_2024} we study the A2744-z7p9OD protocluster. We identify seven new residents of the protocluster, two of which are newly spectroscopically confirmed. We utilise this imaging data of all 23 PRGs to obtain their intrinsic spectral properties, such as the Balmer break and UV slopes. These are seen to be stronger and shallower, respectively, relative to field galaxies at this redshift, evidencing that most of the protocluster residents are already at a more advanced stage of their evolution than is typical. We additionally utilise SED-fitting to infer the stellar masses and star-formation histories of these galaxies. We ultimately identify diverging properties between core and non-core resident galaxies, with the former being composed of massive, dusty galaxies with declining star-formation histories, while the latter are largely characterised as younger galaxies experiencing ongoing bursts of star formation. The damping around the Lyman-break caused by high neutral hydrogen column densities is so extreme that it suppresses the flux in the wide-band F115W filter, and hence we utilise the best-fit SED model to estimate the neutral hydrogen column density of each galaxy. We find extreme column densities around core galaxies (these are normally above $N_{\rm HI}\gtrsim 10^{22}\ {\rm cm^{-2}}$; an order of magnitude above the median column density around high-redshift galaxies from \citealt{mason_constraints_2025}), while non-core galaxies have varying densities (from below $N_{\rm HI}< 10^{21.5}\ {\rm cm^{-2}}$ to $N_{\rm HI}= 10^{23}\ {\rm cm^{-2}}$), suggesting that the neutral hydrogen distribution is in a filamentary structure. 

We identify clear trends in the properties of PRGs compared to typical high-redshift galaxies:
\vspace{-0.2cm}
\begin{itemize}
    \item Extreme neutral hydrogen column densities ($N_{\rm HI}\gtrsim 10^{22}\ {\rm cm^{-2}}$) are observed in the core of the protocluster; an order of magnitude above the typical column densities observed around high-redshift galaxies.
    \item PRGs have redder UV-slopes than is typical in field galaxies ($\Delta\beta\approx 0.4$), evidencing an early divergence in the properties of PRGs.
    \item Core-resident galaxies appear to be undergoing a synchronized decline in their SFHs, with the identification of two (mini)-quenched galaxies resident within the core of the protocluster.
    \item All core-resident galaxies show Balmer breaks (i.e. $B>1$), at odds with $z\sim 8$ field galaxies ($B\sim 0.8$). In contrast, the majority of non-core galaxies show a break strength of $B\lesssim1$, suggesting that the early evolution of PRGs is exacerbated as a function of position.

\end{itemize}

These results indicate that \pc\ is at a more advanced stage of its evolution than would typically be expected from simulations \citep{chiang_galaxy_2017,lim_flamingo_2024}, which ordinarily suggest that protoclusters at $z\sim8$ are characterised by having highly star-forming dense cores, undergoing an inside-out growth phase. By $z<5$ the cores of protoclusters are expected to be the first regions to show evidence of quenching and dense intracluster gas, with more spatially extended star formation \citep{chiang_galaxy_2017,lim_flamingo_2024}. Such properties are already in place in \pc\ at $z\sim8$. 

This advanced evolution of galaxies in extremely overdense regions poses a crucial question for the role of \pc, and protoclusters more generally, in the reionisation process. While this region hosts a significant overdensity of galaxies and hence ionising photons, to what extent these ionising photons can ionise the IGM is highly uncertain. The high neutral hydrogen column densities and significant dust content of PRGs inhibits the escape of these ionising photons, potentially counteracting their high ionising photon production rate density. However, the recent detection of a strong \Lya-emitting galaxies resident in this protocluster \citep{chen_jwst_2024,cameron_nebular_2024}, ZD4, provides an early indication that the vast number of ionising sources may well outweigh the decreased fraction of escaping ionising photons in this extremely overdense environment.

\begin{acknowledgements}
The work presented in this paper is based on observations made with the NASA/ESA/CSA James Webb Space Telescope. The data were obtained from the Mikulski Archive for Space Telescopes at the Space Telescope Science Institute, which is operated by the Association of Universities for Research in Astronomy, Inc., under NASA contract NAS 5-03127 for JWST. These observations are associated with program 2561 and 4111.

This work has received funding from the Swiss State Secretariat for Education, Research and Innovation (SERI) under contract number MB22.00072, as well as from the Swiss National Science Foundation (SNSF) through project grant 200020\_207349.
The Cosmic Dawn Center (DAWN) is funded by the Danish National Research Foundation under grant DNRF140.
L.C. and C.C. acknowledge support from the French government under the France 2030 investment plan, as part of the Initiative d’Excellence d’Aix-Marseille Université – A*MIDEX AMX-22-RE-AB-101 and partial supported by the “PHC GERMAINE DE STAEL” programme (project number: 52217VG), funded by the French Ministry for Europe and Foreign Affairs, the French Ministry for Higher Education and Research, the Swiss Academy of Technical Sciences (SATW) and the State Secretariat for Education, Research and Innovation (SERI). DS acknowledges support from the STFC, grant code ST/W000997/1. WM thanks the Science and Technology Facilities Council (STFC) Center for Doctoral Training (CDT) in Data Intensive Science at the University of Cambridge (STFC grant number 2742968) for a PhD studentship. CS acknowledges support from the Science and Technology Facilities Council (STFC), by the ERC through Advanced Grant 695671 ``QUENCH'', by the UKRI Frontier Research grant RISEandFALL. YF acknowledges supports from JSPS KAKENHI Grant Numbers JP22K21349 and JP23K13149. ALD thanks the University of Cambridge Harding Distinguished Postgraduate Scholars Programme and Technology Facilities Council (STFC) Center for Doctoral Training (CDT) in Data intensive science at the University of Cambridge (STFC grant number 2742605) for a PhD studentship, and acknowledges support by the Royal Society Research Grant G125142.
\end{acknowledgements}

\bibliographystyle{aa}
\bibliography{references_downloaded} 

\begin{appendix}

\section{Spectroscopic confirmation of two new candidates}
\label{app:spec}
\begin{figure}
    \centering
    \includegraphics[width=0.99\linewidth]{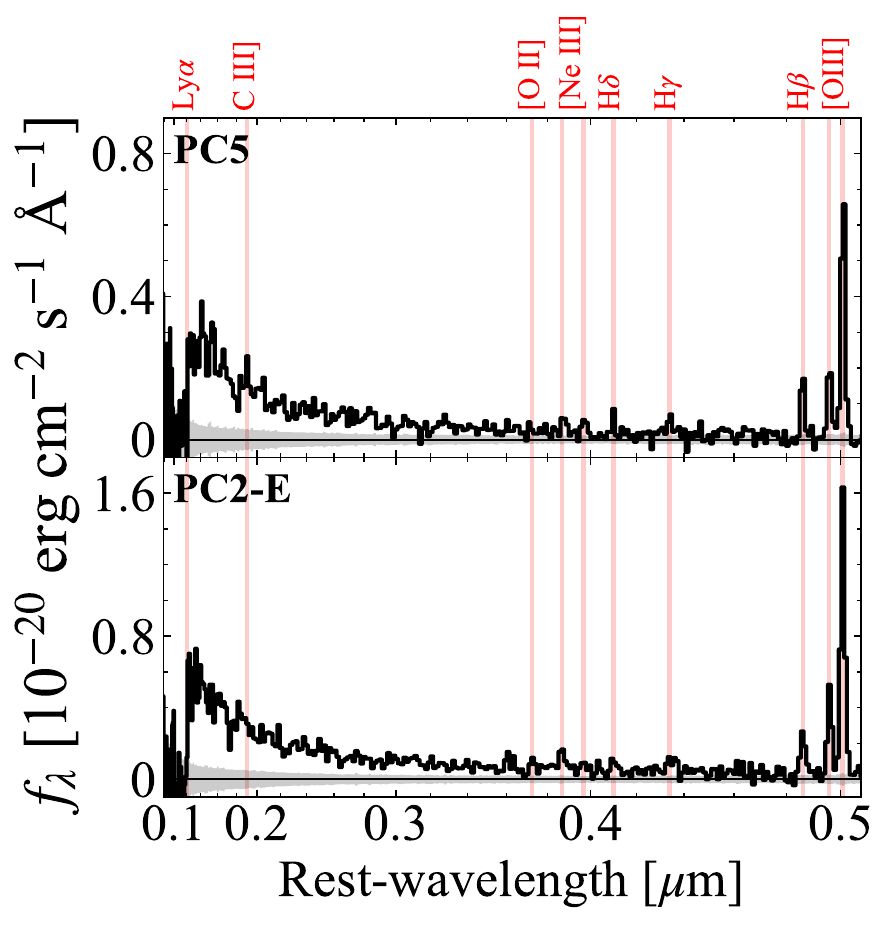}
    \caption{The spectra of PC5 (top) and PC2-E (bottom). The red shaded lines correspond to the expected position of some of the typical emission lines observed from high-redshift galaxies. Both spectra show strong \OIII and \Hb emission lines that facilitate the measurement of their redshift.}
    \label{fig:PC5+PC2-E}
\end{figure}

The spectra of PC5 and PC2-E have recently been observed as part of the UNCOVER program \cite{bezanson_jwst_2024}, during a repeat visit (2561:6:2). In Figure~\ref{fig:PC5+PC2-E} we show the Dawn JWST Archive (DJA) reduction of these objects, which utilises the custom-made pipeline \texttt{msaexp} \citep{brammer_msaexp_2023}\footnote{https://github.com/gbrammer/msaexp}. Further details of the reduction pipeline are given in \cite{de_graaff_rubies_2024,heintz_jwst-primal_2025}. These spectra show strong \OIII and \Hb emission lines, and by fitting these with Gaussians we obtain the spectroscopic redshift of these sources. Again we utilise a MC propagation technique to estimate the uncertainties in this redshift measurement. These redshifts are reported in Table~\ref{tab:Targets}.

\section{SEDs and SFHs}
\begin{figure}
    \centering
    \includegraphics[width=0.99\linewidth]{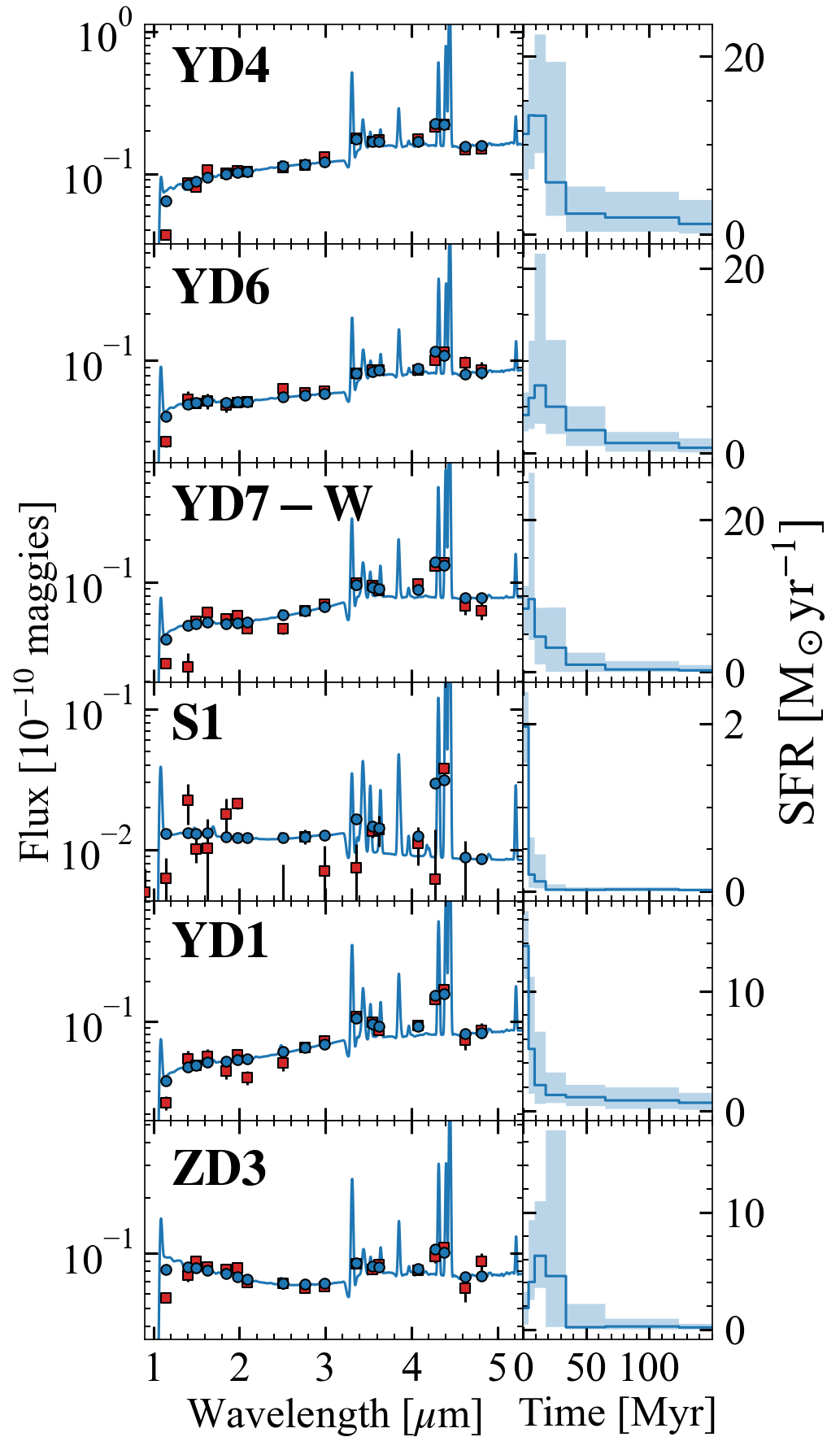}
    \caption{(left:) The observed and best fit SED of PRGs (where we exclude filters where the SED is not detected, $\lambda < 1 \mu$m) and (right:) their associated best-fit SFHs. See Figure~\ref{fig:(mini)-quenched} for further details.}
    \label{fig:SED+SFH1}
\end{figure}

\begin{figure}
    \centering
    \includegraphics[width=0.99\linewidth]{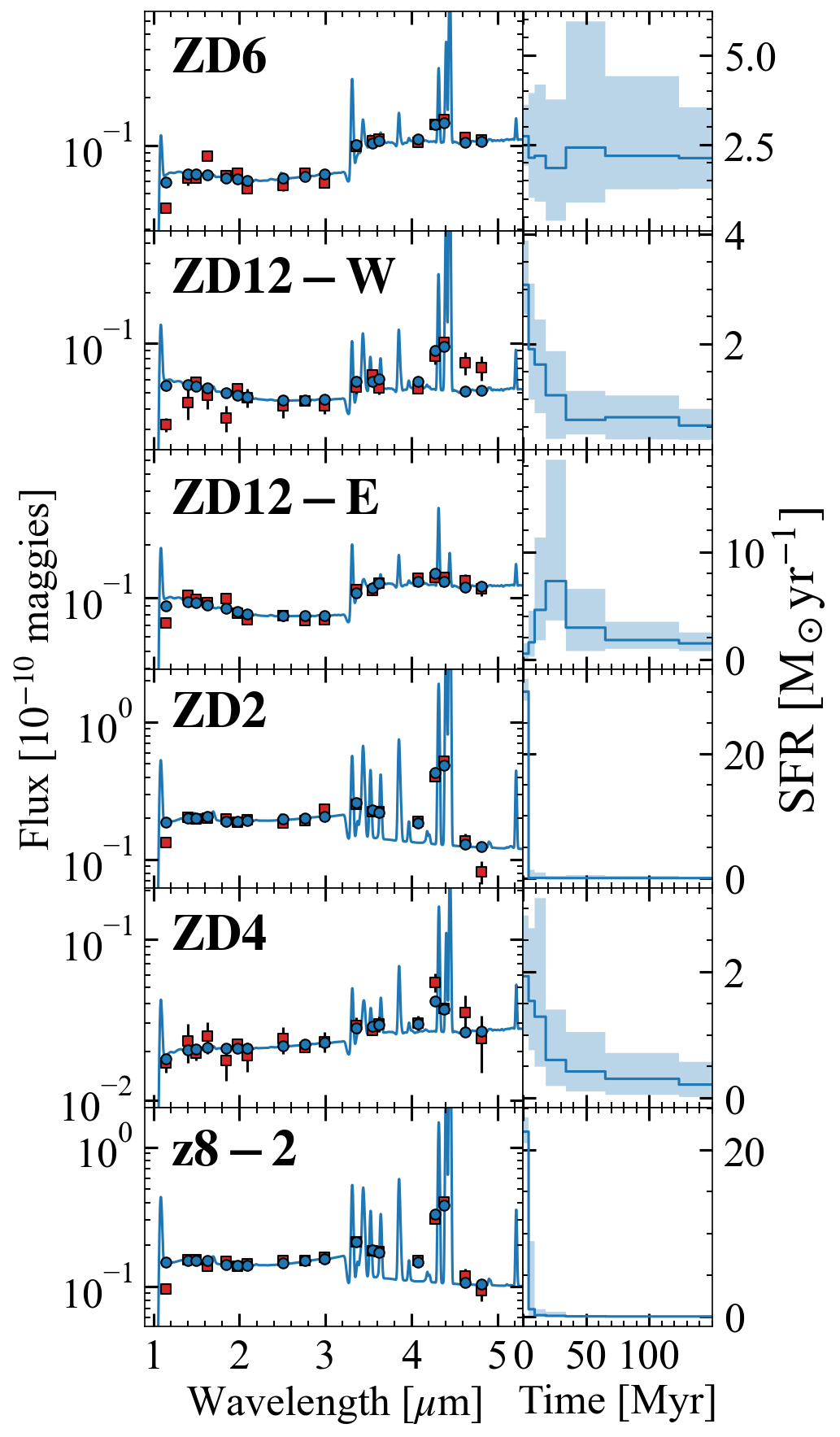}
    \caption{Same as Figure~\ref{fig:SED+SFH1}.}
    \label{fig:SED+SFH2}
\end{figure}

\begin{figure}
    \centering
    \includegraphics[width=0.99\linewidth]{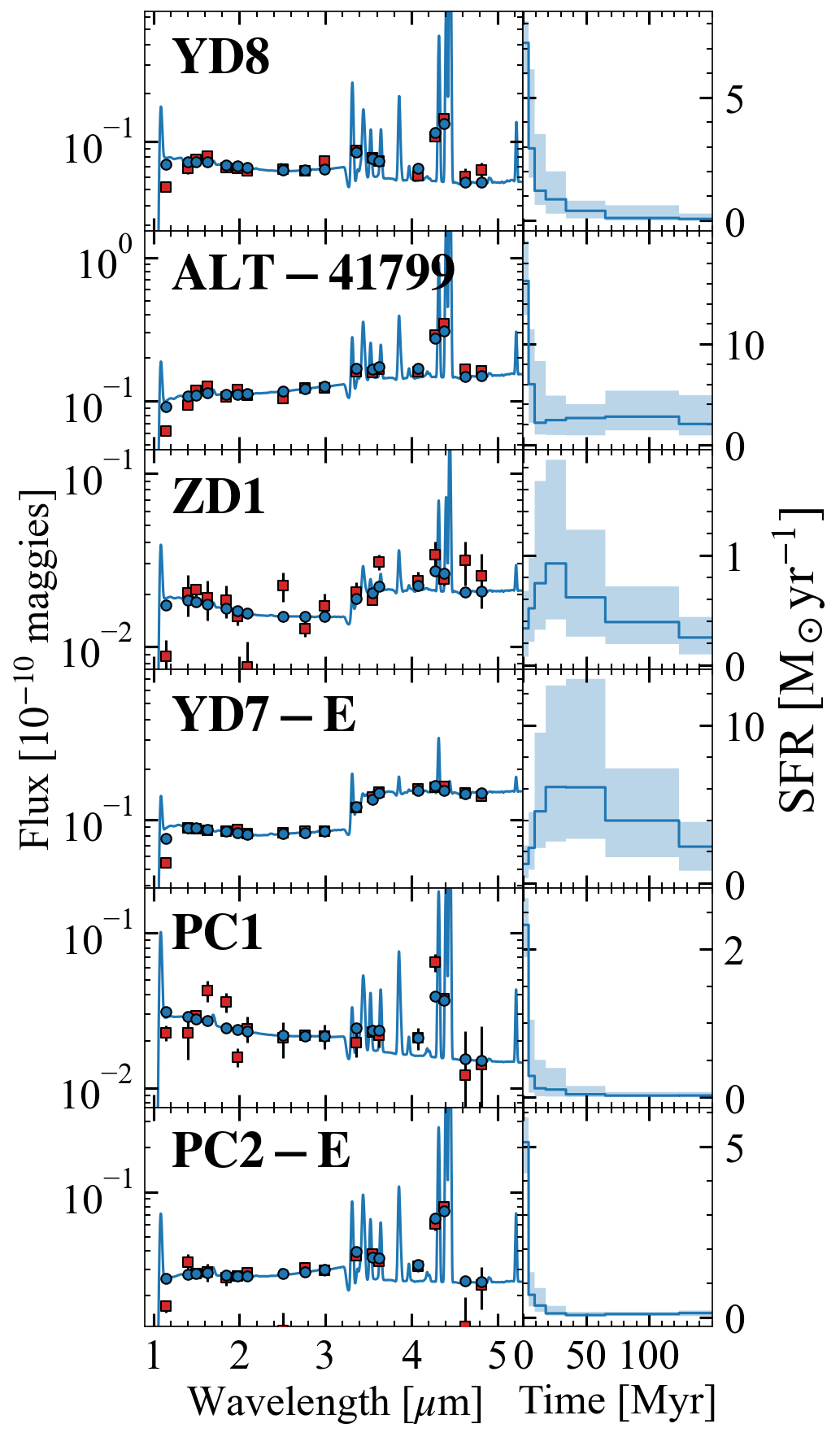}
    \caption{Same as Figure~\ref{fig:SED+SFH1}.}
    \label{fig:SED+SFH3}
\end{figure}

\begin{figure}
    \centering
    \includegraphics[width=0.99\linewidth]{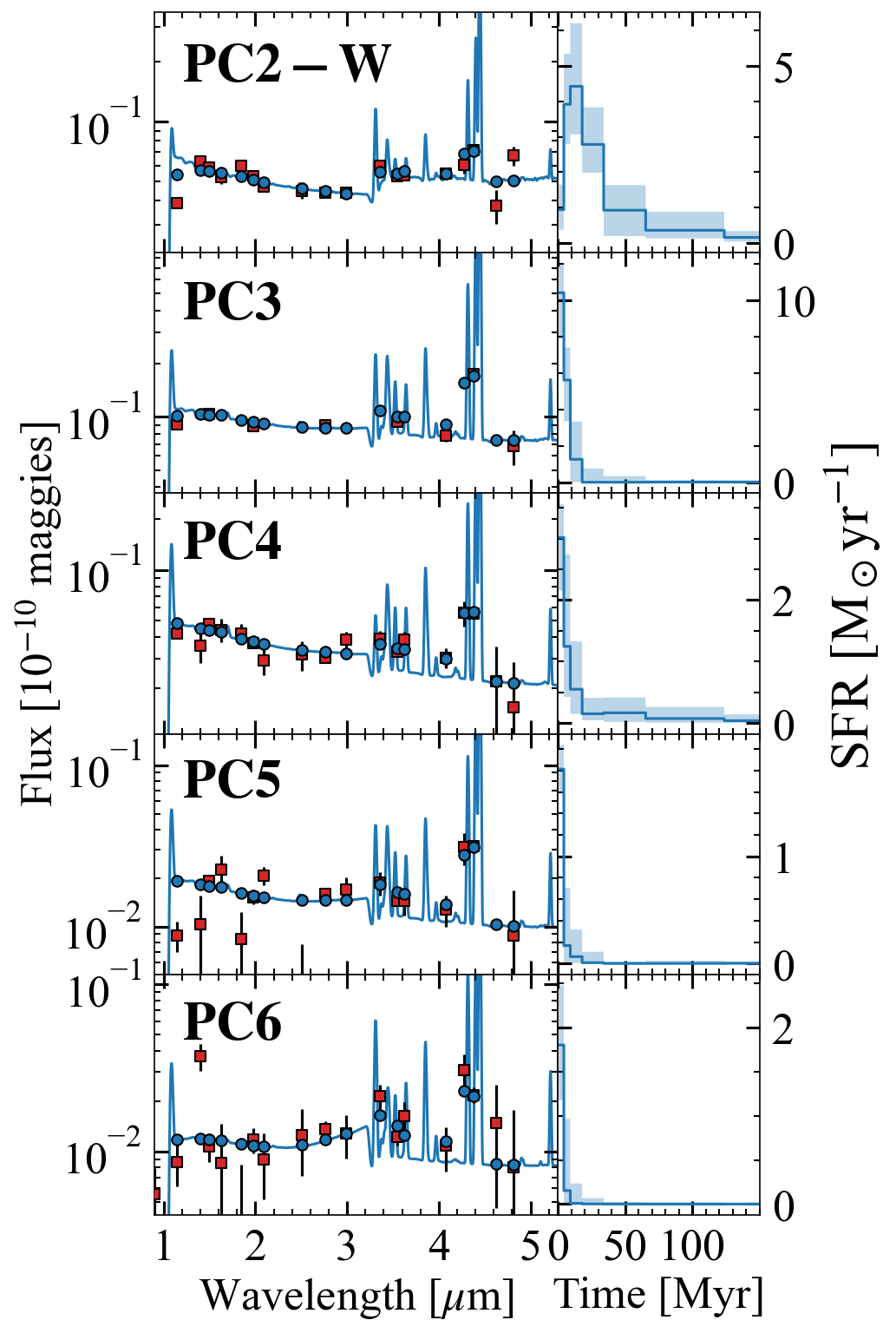}
    \caption{Same as Figure~\ref{fig:SED+SFH1}.}
    \label{fig:SED+SFH4}
\end{figure}

In the following section we include the photometric data and best-fit SEDs from our \texttt{PROSPECTOR} (discussed in more detail in Section~\ref{sec:sed-fitting}) for our sample of PRGs in Figure~\ref{fig:SED+SFH1}, ~\ref{fig:SED+SFH2}, ~\ref{fig:SED+SFH3} and ~\ref{fig:SED+SFH4}. We additionally show the SFHs returned from this SED-fitting. These galaxies range from those with strong emission lines and Balmer jumps, characterised by a recent, very strong burst in their SFHs, to weak emission line sources with strong Balmer breaks, characterised by elongated SFHs that are declining at the time of observation. 


\end{appendix}

\end{document}